\documentclass[fleqn,usenatbib]{mnras}
\DeclareSymbolFont{cmletters}{OML}{cmm}{m}{it}
\DeclareMathSymbol{v}{\mathalpha}{cmletters}{"76}

\usepackage{ae,aecompl}
\usepackage{times}
\usepackage{soul}

\usepackage{tabularx,ragged2e,booktabs,caption}
\usepackage{amsmath}
\usepackage{amssymb}
\usepackage{epsfig}
\usepackage{graphicx}
\usepackage{ifthen}
\usepackage{latexsym}
\usepackage{rotating}
\usepackage{times,epsf}
\usepackage{txfonts}
\usepackage{varioref}
\usepackage{verbatim}
\usepackage{array}
\usepackage{url}
\usepackage{color}
\usepackage[T1]{fontenc}
\usepackage{epstopdf}
\usepackage{ulem}

\def\gsim{\mathrel{\raise.5ex\hbox{$>$}\mkern-14mu
             \lower0.6ex\hbox{$\,\sim$}}}
\def\lsim{\mathrel{\raise.3ex\hbox{$<$}\mkern-14mu
             \lower0.6ex\hbox{$\,\sim$}}}

\newcommand{\be}{\begin{equation}}
\newcommand{\ee}{\end{equation}}
\newcommand{\bea}{\begin{eqnarray}}
\newcommand{\eea}{\end{eqnarray}}

\title[Rayleigh-Taylor Instability II]{The magnetic Rayleigh-Taylor instability around astrophysical black holes}
\author[D. B. Papadopoulos, I. Contopoulos]
       {D. B. Papadopoulos$^{1}$, I. Contopoulos$^{2,3}$\thanks{E-mail: icontop@academyofathens.gr}\\
$^1$ Physics Department, Aristotle University of Thessaloniki, Thessaloniki 54124, Greece\\
$^2$ Research Center for Astronomy and Applied Mathematics, Academy of Athens, Athens 11527, Greece\\
$^3$ National Research Nuclear University (MEPhI), Moscow 115409, Russia\\
}

\begin{document}

\maketitle

\label{firstpage}

\begin{abstract}
We investigate the development of the magnetic Rayleigh-Taylor instability at the inner edge of an astrophysical disk around a spinning central black hole. We solve the equations of general relativity that govern small amplitude oscillations of a discontinuous interface in a Keplerian disk threaded by an ordered magnetic field, and we derive a stability criterion that depends on the central black hole spin and the accumulated magnetic field. We also compare our results with the results of GR MHD simulations of black hole accretion flows that reach a magnetically arrested state (MAD). We found that the instability growth timescales that correspond to the simulation parameters are comparable to the corresponding timescales for free-fall accretion from the ISCO onto the black hole. We thus propose that the Rayleigh-Taylor instability disrupts the accumulation of magnetic flux onto the black hole horizon as the disk reaches a MAD state.
\end{abstract}


\section{Introduction}
Magnetic fields are believed to play a fundamental role in powering energetic astrophysical sources such as active galactic nuclei, X-ray binaries and gamma-ray bursts. Extensive theoretical research over the past four decades has most convincingly shown that magnetic fields contribute to the extraction of rotational energy from spinning astrophysical black holes as proposed forty years ago by \citet{BZ77}. The fundamental parameter that determines the efficiency of this process is the amount of magnetic flux $\Phi_{\rm BH}$ that threads the black hole horizon. It is well known that magnetized accretion may bring magnetic flux toward the black hole, but when matter finally crosses the horizon, the magnetic field decouples from the matter and leaves the black hole at light-crossing times, unless there is some external medium preventing it from escaping to infinity. In astrophysical black holes, the role of this external medium is played by the surrounding accretion disk.

This configuration may be naively described as the `heavy' disk material holding the `light' magnetic field from escaping `buoyantly', as in water over oil in vertical gravitational equilibrium. We are thus very much interested in studying the development of the magnetic Rayleigh-Taylor (hereafter RT) instability around the inner edge of the accretion disk when a large scale vertical magnetic field is present inside it. This will help us understand what limits the maximum amount of magnetic flux that threads the black hole horizon which, as we said above, is the fundamental parameter that characterizes the efficiency of the Blandford-Znajek process.

In a previous work \cite[][hereafter Paper~I]{CKP16}, we investigated the magnetic RT instability in a non-rotating equatorial disk of plasma at the position of the innermost stable circular orbit (hereafter ISCO) around a slowly rotating black hole. We obtained very low limits for the maximum flux that can be stably held inside the ISCO. We found that a disk around a Schwarzschild black hole is unstable, and that black hole rotation slightly stabilizes the system. On the contrary, in their simulations of magnetized black hole accretion, \cite{TMN12} observed only a mild dependence of the accumulated dimensionless magnetic flux on the black hole spin. In Paper~I, we speculated that this may be due to our neglect of rotation and/or the dynamics of accretion. We decided to extend our analysis and consider a Keplerian isothermal incompressible equatorial disk around a Kerr black hole. The matter distribution extends practically down to the ISCO, inside which free-fall accretion abruptly reduces the matter density $\rho$ \citep[e.g.][]{Petal10}. There is also a vertical magnetic field $B$ that accumulates inside it. We will thus treat the ISCO as a discontinuous interface along which we will investigate the development of the magnetic RT instability. In \S~2 we establish the general theoretical problem, and apply it to the study of a Kerr black hole surrounded by a Keplerian disk. In \S~3 we obtain the main equation that yields the stability criterion and the instability growth timescale as a function of the accumulated magnetic field and the spin of the central black hole. Finally, in \S~4 we discuss the astrophysical implications of our results.

\section{Small perturbation analysis}

\subsection{General relativistic MHD in 3+1 formalism}

As in Paper~I, we follow here the 3+1 (space+time) formalism of general relativistic magnetohydrodynamics (GRMHD) developed by \cite{a1}. In this paper we will work in geometrical units in which $c=G=1$. We introduce spatial magnetic and electric fields ${\bf B}$ and ${\bf E}$ respectively measured by fiducial observers with 4-velocity $U^{\mu}$. In that formalism, Maxwell's equations
$F_{;\beta}^{\alpha\beta}=4\pi J^{\alpha}$,
$F_{[\alpha\beta;\gamma]}=0$, and $J_{;\alpha}^{\alpha}=0$ yield
\begin{eqnarray}\label{k9}
&&\tilde{\nabla}\cdot \tilde{E} = 4\pi\rho_e\nonumber\\
&&\tilde{\nabla}\cdot \tilde{B} = 0\nonumber\\
&&D_{\tau}\tilde{E}+\frac{2}{3}\theta
\tilde{E}-\tilde{\sigma}\cdot
\tilde{E} = \frac{1}{\alpha}\tilde{\nabla}\times (\alpha \tilde{B})-4\pi \tilde{J}\nonumber\\
&&D_{\tau}\tilde{B}+\frac{2}{3}\theta
\tilde{B}-\tilde{\sigma}\cdot \tilde{B} =
-\frac{1}{\alpha}\tilde{\nabla}\times (\alpha \tilde{E})\label{k9}
\end{eqnarray}
Here, $D_{\tau} M^{\beta}\equiv M^{\beta}~_{;\mu}
U^{\mu}-U^{\beta}a_{\mu} M^{\mu}$ is the Fermi derivative,
$\theta$ and $\tilde{\sigma}$ are the expansion and shear of the spacetime metric respectively, and $\rho_e$ is the electric charge density in the rest frame of the fluid.
$\alpha\equiv d\tau/dt$ is the `lapse function', where $\tau$ is the fluid proper time, and $t$ is the global coordinate time.

The evolution of the magnetized fluid is characterized by the divergence of the total stress-energy tensor $T^{\mu\nu}\equiv T_{\rm
matter}^{\mu\nu}+T_{\rm EM}^{\mu\nu}$ , namely
\begin{eqnarray}\label{k11}
&&T^{\mu\nu}_{;\nu}=0\ .
\end{eqnarray}
This yields
\begin{eqnarray}
&&D_{\tau}\varepsilon+\theta\varepsilon+\frac{1}
{\alpha^2}\tilde{\nabla}\cdot(\alpha^2 \tilde{S})
+W^{jk}(\sigma_{jk}+\frac{1}{3}\theta\gamma_{jk})=-\tilde{J}\cdot
\tilde{E}\nonumber\\
&&D_{\tau}\tilde{S}+\frac{4}{3}\theta
\tilde{S}+\tilde{\sigma}\cdot\tilde{S}+\varepsilon
\tilde{a}+\frac{1}{\alpha}\tilde{\nabla}\cdot(\alpha
\tilde{W})= \nonumber\\
&&\rho_e\tilde{E}+\tilde{J}\times\tilde{B}\label{k12}
\end{eqnarray}
where,
\begin{eqnarray}\label{k13}
&&\varepsilon\equiv T_{\rm matter}^{\mu\nu}U_{\mu}U_{\nu}\nonumber\\
&&S^{\alpha}\equiv \gamma^{\alpha}~_{\mu} T_{\rm matter}^{\mu\nu}
U_{\nu}\nonumber\\
&&W^{\alpha\beta}\equiv \gamma^{\alpha}~_{\mu} T_{\rm matter}^{\mu\nu} \gamma^{\beta}~_{\nu}\nonumber\\
&&\theta\equiv U^{\mu}_{;\mu},~~ a^{\mu}\equiv
U^{\mu}_{;\nu}U^{\nu}
, ~~\nonumber\\
&&\sigma_{ab}\equiv \frac{1}{2}\gamma^{\mu}~_{a}\gamma^{\nu}~_{b}(U_{\mu;\nu}+U_{\nu;\mu})-\frac{1}{3}\theta \gamma_{ab}\nonumber\\
&&\tilde{L}\cdot \tilde{M}=\gamma^{ij} L_i
M_j,~~(\tilde{L}\times\tilde{M})^j=\epsilon^{ijk} L_j M_{\rm K}\ .
\end{eqnarray}
$\gamma^{\alpha\beta}=g^{\alpha\beta}+U^\alpha U^\beta$ is
the projection tensor, and $\alpha$ is the lapse function. Latin
indices take values $1,2,3$ and Greek ones $0,1,2,3$. Vectors and
tensors with tilde are purely spatial. For an ideal fluid with
rest energy density $\rho$, 3-velocity $\tilde{v}$, and pressure $p$ we have
\begin{eqnarray}\label{k14}
&&\Gamma=(1-\tilde{v}^2)^{-1/2},~~\varepsilon= \Gamma^2(\rho+p\tilde{v}^2)\nonumber\\
&&\tilde{S}=(\rho+p)\Gamma^2
\tilde{v},~~\tilde{W}=(\rho+p)\Gamma^2\tilde{v}\otimes\tilde{v}+p\tilde{\gamma}\
.
\end{eqnarray}
Moreover, we assume conservation of mass (or equivalently baryon number) in the flow, namely
\begin{eqnarray}
&&(\rho u^{\mu})_{;\mu}=0
\label{baryonnumber0}
\end{eqnarray}
\citep{xn1}. In Appendix~A we show that, in 3+1 formalism,  eq.~(\ref{baryonnumber0}) can be rewritten as
\begin{eqnarray}
&&D_{\tau}(\Gamma\rho)+\Gamma\rho\theta+\tilde{\nabla}\cdot (\Gamma\rho \tilde{\upsilon})=0\ .
\label{baryonnumber}
\end{eqnarray}
We further assume an equation of state $p\equiv p(\rho)$ from which one can
deduce the `speed of sound' $c_s\equiv (dp/d\rho)^{1/2}$. For simplicity, in what follows we will consider only an isothermal fluid with
\begin{eqnarray}
\label{isothermal}
&&\frac{p}{\rho}=c_s^{2}=\mbox{const.}
\end{eqnarray}
Finally, we assume ideal MHD conditions, namely
\begin{eqnarray}\label{l1x}
&&\tilde{E}=-\tilde{v}\times \tilde{B}
\end{eqnarray}
As in Paper~I, we investigate the development of the magnetic RT instability in the astrophysical context of a thin accretion disk, thus we restrict our analysis to the equatorial plane ($\theta=\pi/2$). This time the disk is not stationary, but is in Keplerian rotation around the central black hole.

\subsection{Kerr spacetime}

In Boyer-Lindquist coordinates the Kerr metric reads
\begin{eqnarray}\label{k1x}
ds^2 & = & g_{tt} dt^2+2g_{t\phi}dt d\phi +g_{rr}
dr^2+g_{\theta\theta} d\theta^2+g_{\phi\phi} d\phi^2\nonumber\\
& = & -(1-\frac{2M r}{\Sigma})dt^2-\frac{4 M a
r\sin^2{\theta}}{\Sigma}dt d\phi\nonumber\\
&&+\frac{\Sigma}{\Delta} dr^2+\Sigma
d\theta^2+\frac{A}{\Sigma}\sin^2{\theta}d\phi^2\end{eqnarray}
where $M$ is the mass of the black hole, $a$ is the angular
momentum per unit mass $(0\leq a \leq M)$, and
\begin{eqnarray}\label{k2x}
&&\Delta \equiv r^2-2M r+a^2\nonumber\\
&&\Sigma \equiv r^2+a^2\cos^2{\theta}\nonumber\\
&&A \equiv(r^2+a^2)^2-a^2\Delta\sin^2{\theta}\end{eqnarray}
(Cowling~1941).

For our further study  we need the components of the 4-velocity of fiducial observers, now identified as ZAMOs (Zero Angular Momentum Observers), namely
\begin{equation}\label{k3x}
U^{\mu}=(\frac{1}{\alpha},0,0,\frac{\omega}{\alpha})\ ,\ \
U_{\mu}=(-\alpha,0,0,0)
\end{equation}
where
\begin{equation}\label{k5x}
\alpha=\sqrt{\frac{\Delta\Sigma}{A}},~~\omega=\frac{2M
ar}{A}\end{equation}
In the Kerr spacetime with 4-velocity given
by eq.~(\ref{k3x}), the expansion $\theta$ vanishes, the shear $\tilde{\sigma}$
has two non-zero components $\sigma^{13}$ and $\sigma^{23}$, and
$\sigma_{\alpha\beta}\gamma^{\alpha \beta}=0$ \citep[][eq.~2.5]{a1}. The acceleration $a^{\mu}$ is
given by
\begin{eqnarray}
\label{k6ax}
a^{\mu}&=&\frac{Ma^2 r^4}{\Sigma^2 A}\left(0,\right.\nonumber\\
&&-\cos^2{\theta}[(1+(\frac{a}{r})^2)^2-\frac{4M}{r}]
+\ (\frac{r}{a})^2 [(1+(\frac{a}{r})^2)^2-\frac{4M}{r}]
\ ,\nonumber\\
&&
\left.(1+(\frac{a}{r})^2)\frac{\sin{2\theta}}{r},0\right)
\end{eqnarray}
$\gamma_{ij}$ is the spatial metric on the space-like hypersurface
$x^0\equiv t=$~const., with normal vector $n_{\alpha}$
\begin{equation}\label{k8x}
n_{\alpha}=(-\alpha,0,0,0),~~n^{\alpha}=\frac{1}{\alpha}(1,-\beta^1,-\beta^2,-\beta^3)\end{equation}
where $\beta^i=\gamma^{ij}g_{0j}$.

\subsection{Perturbed equations}

We consider only small perturbations of physical quantities $f$ as
\begin{equation}
\label{per1}
f(t,\tilde{r})=f(\tilde{r})+\delta\! f(t,\tilde{r})
\end{equation}
%
where the perturbations in the equatorial plane $\theta=\pi/2$ are of the form
\begin{equation}\label{per1a}
\delta\! f(t,r,\frac{\pi}{2},\phi)=\delta\! f(r)\, e^{nt+i m\phi}
\end{equation}
and keep only linear terms of the perturbations. In this case, in the Cowling approximation of a fixed Kerr spacetime,
the zeroth order MHD equations are:
\begin{eqnarray}\label{x1a}
&\tilde{\nabla}\cdot \tilde{E}=4\pi\rho_e,~~\tilde{\nabla}\cdot \tilde{B}=0\nonumber\\
&D_{\tau}\tilde{E}-\tilde{\sigma}\cdot \tilde{E}=\tilde{\nabla}\times \tilde{B}-\tilde{B}\times\tilde{a}-4\pi \tilde{J}\nonumber\\
&D_{\tau}\tilde{B}-\tilde{\sigma}\cdot \tilde{B}=-\tilde{\nabla}\times \tilde{E}+\tilde{E}\times\tilde{a}\nonumber\\
&D_{\tau}\rho_e+\tilde{\nabla}\cdot \tilde{J}+\tilde{J}\cdot\tilde{a}=0\nonumber\\
&D_{\tau}\varepsilon+\tilde{\nabla}\cdot \tilde{S}+2\tilde{S}\cdot\tilde{a}+W^{jk}\sigma_{jk}=-\tilde{J}\cdot\tilde{E}\nonumber\\
&D_{\tau}\tilde{S}+\tilde{\sigma}\cdot \tilde{S}+\varepsilon\tilde{a}+\tilde{\nabla}\cdot \tilde{W}+\tilde{W}\cdot\tilde{a}=(\rho_e\tilde{E}+\tilde{J}\times \tilde{B})
\end{eqnarray}
The first order MHD equations are:
\begin{eqnarray}\label{mhd1a}
&\tilde{\nabla}\cdot \delta\tilde{E}=4\pi\delta\rho_e\\\label{mhd1b}
&\tilde{\nabla}\cdot \delta\tilde{B}=0\\
&\label{mhd1c}
D_{\tau}\delta\tilde{E}=\tilde{\nabla}\times \delta\tilde{B}+\tilde{a}\times \delta \tilde{B}+\tilde{\sigma}\cdot \delta\tilde{E}-4\pi \delta\tilde{J}\\
&\label{mhd1d}
D_{\tau}\delta \tilde{B}=-\tilde{\nabla}\times \delta\tilde{E}-\tilde{a}\times \delta\tilde{E}+\tilde{\sigma}\cdot\delta\tilde{B}\\
&\label{mhd1e}
D_{\tau}\delta\rho_e+\delta\tilde{J}\cdot \tilde{a}+\tilde{\nabla}\cdot \delta\tilde{J}=0\\
&\label{mhd2a}
D_{\tau}\delta\varepsilon +2\delta\tilde{S}\cdot \tilde{a}+\tilde{\nabla}\cdot \delta\tilde{S}+\tilde{\sigma}\cdot\delta \tilde{W}
=-\delta\tilde{J}\cdot\tilde{E}-\tilde{J}\cdot\delta\tilde{E}\\
&
\label{mhd2b}
D_{\tau}\delta\tilde{S}+\tilde{a}\delta\varepsilon +\delta\tilde{W}\cdot \tilde{a}+\tilde{\nabla}\cdot \delta\tilde{W}+\tilde{\sigma}\cdot\delta\tilde{S}\nonumber\\
&=(\delta \rho_e\tilde{E}+\delta\tilde{J}\times
\tilde{B})+(\rho_{e}\delta\tilde{E}+\tilde{J}\times \delta\tilde{B})\label{s1t}
\end{eqnarray}
The perturbed eqs.~(\ref{k14}) become
\begin{eqnarray}\label{vp2}
& \delta \tilde{\upsilon}^2 = 2\tilde{\upsilon}\, \tilde{\delta\upsilon}\ \ ,\ \delta \Gamma^2 = 2\tilde{\upsilon}\, \delta \tilde{\upsilon}\,(1-\tilde{\upsilon}^2)^{-2}\nonumber\\
&\delta\varepsilon = \delta\Gamma^2 (\rho+p\tilde{\upsilon}^2)+\Gamma^2(\delta\rho+\delta p \tilde{\upsilon}^2+p\delta\tilde{\upsilon}^2)\nonumber\\
&\delta \tilde{S} = [(\delta\rho+\delta p)\Gamma^2+(\rho+p)\delta\Gamma^2]\tilde{\upsilon}+(\rho+p)\Gamma^2\delta\tilde{\upsilon}\nonumber\\
&\delta \tilde{W} = [(\delta\rho+\delta p)\Gamma^2+(\rho+p)\delta\Gamma^2]\tilde{\upsilon}\otimes\tilde{\upsilon}\nonumber\\
&+(\rho+p)\Gamma^2[\delta\tilde{\upsilon}\otimes\tilde{\upsilon}+\tilde{\upsilon}\otimes\delta\tilde{\upsilon}]+\tilde{\gamma}\delta p
\end{eqnarray}
and the ideal MHD condition (eq.~\ref{l1x}) yields
\begin{equation}\label{s11x}
\delta\tilde{E}=-\delta\tilde{\upsilon}\times \tilde{B}-\tilde{\upsilon}\times \delta\tilde{B}
\end{equation}

\subsection{Keplerian Disc}

For our further study we will assume a simple Keplerian flow configuration in the equatorial plane, namely
\begin{equation}
\tilde{\upsilon}=(0,0,\upsilon^\phi)\ ,
\end{equation}
where, because of symmetry,
\begin{equation}\label{v3}
J^{\mu}=(0,0,0,J^{\phi})\ \mbox{and}\ \tilde{B}=(0,B^{\theta},0)\ .
\end{equation}
We have set $\upsilon^r=0$, but allow for nonzero $\delta\upsilon^r$ equatorial velocity perturbations. We do not consider off-plane perturbations\footnote{Off-plane perturbations along an initially vertical magnetic field may be related to the well studied magnetorotational instability \citep[MRI;][]{BH91}. Here, we are interested only in the development of the Rayleigh-Taylor instability triggered by equatorial motions.} and set $\delta v^\theta=0$. Furthermore, we will assume for simplicity that
$\rho_e=0$, $\delta\rho_e=0$, and that the fluid is incompressible such that
\begin{equation}\label{v4}
\tilde{\nabla}\cdot\tilde{\upsilon}=0~\mbox{~(zeroth order)~},
\tilde{\nabla}\cdot\delta\tilde{\upsilon}=0~~\mbox{~(first order)~}.
\end{equation}
The latter yields
\begin{equation}\label{v4b}
\delta \upsilon^{\phi}=-\frac{(r^2\delta \upsilon^r)_{,r}}{i m r^2}
\end{equation}
The perturbed eq.~(\ref{baryonnumber}) yields (see Appendix~A)
\begin{eqnarray}\label{vv5}
&&D_{\tau}\delta \rho+\delta\tilde{\upsilon}\cdot\tilde{\nabla}\rho+\tilde{\upsilon}\cdot\tilde{\nabla}\delta\rho\nonumber\\
&&=-\rho\delta\upsilon^r\partial_r(\tilde{\upsilon}^2)-\frac{\rho}{2(1-\tilde{\upsilon}^2)}[\upsilon^\phi\partial_\phi(\delta\tilde{\upsilon}^2)]
\end{eqnarray}

We now re-write eq.~(\ref{mhd2b}) as
\begin{eqnarray}\label{v5}
\tilde{\nabla}\cdot \delta\tilde{W}&=&-D_{\tau}\delta\tilde{S}-\tilde{a}\delta\varepsilon -\delta\tilde{W}\cdot \tilde{a}-\tilde{\sigma}\cdot\delta\tilde{S}\nonumber\\
&&+\ \delta\tilde{J}\times \tilde{B}+\tilde{J}\times \delta\tilde{B}
\end{eqnarray}
We split $\delta\tilde{W}=\delta\hat{W}+\tilde{\gamma}\delta p$, where $\delta\tilde{W}$ has contravariant components
\begin{eqnarray}\label{v7}
\delta \hat{W}^{ij}&=&f_1\upsilon^i\upsilon^j+f_2(\upsilon^i\delta\upsilon^j+\upsilon^j\delta\upsilon^i),~~\mbox{~with~}\nonumber\\
f_1&=&(1+c_s^2)(\Gamma^2\delta\rho+\rho\delta\Gamma^2),~~\mbox{and~}\nonumber\\
f_2&=&(1+c_s^2)\rho\Gamma^2
\end{eqnarray}
Recall that
\begin{equation}\label{v8}
\tilde{\nabla}\cdot \delta \tilde{W} =\frac{1}{\sqrt{\gamma}}\frac{\partial}{\partial x^j}[\sqrt{\gamma}\delta \hat{W}^{ij}]+\Gamma^i_{kl}\delta \hat{W}^{kl}+\gamma^{ij}\delta p_{,j}
\end{equation}
where $\gamma=\det(\gamma_{ij})=\Sigma^2 /\alpha^{2}$ is the determinant of the radial 3-metric of the Kerr space time. Next, eqs.~(\ref{v5}) \& (\ref{v8}) give
\begin{eqnarray}\label{d1}
\gamma^{ij}\delta p_{,j}&=&-\frac{1}{\sqrt{\gamma}}\frac{\partial}{\partial x^j}[\sqrt{\gamma}\delta \hat{W}^{ij}]-\Gamma^i_{kl}\delta \hat{W}^{kl}\nonumber\\
&&-D_{\tau}\delta S^i-a^i\delta \varepsilon-\delta \hat{W}^{ik}\gamma_{kl}a^l-\sigma^{ik}\gamma_{kl}\delta S^l\nonumber\\
&&+(\delta J\times B)^i+(J\times\delta B)^i,
\end{eqnarray}
where $i=r,\phi$. This yields
\begin{eqnarray}\label{v9}
\gamma^{\phi\phi}\delta p_{,\phi}&=&-f_{1,r}(\upsilon^r\upsilon^{\phi})-f_1(\upsilon^r\upsilon^{\phi})_{,r}-f_{2,r}(\upsilon^r\delta\upsilon^{\phi}+\upsilon^{\phi}\delta\upsilon^r)\nonumber\\
&&-f_2[(\upsilon^r\delta\upsilon^{\phi})_{,r}+(\upsilon^{\phi}\delta\upsilon^r)_{,r}]-(i m)\upsilon^{\phi}[f_1\upsilon^{\phi}+2f_2\delta\upsilon^{\phi}]\nonumber\\
&&-(\frac{2}{r}+\Gamma_{r\phi}^{\phi})[f_1\upsilon^r\upsilon^{\phi}+f_2(\upsilon^r\delta\upsilon^{\phi}+\upsilon^{\phi}\delta\upsilon^r)]\nonumber\\
&&-(\frac{1}{\alpha})[\Gamma_{t r}^{\phi}+\omega\Gamma_{\phi r}^{\phi}-\omega a_r+\gamma_{rr}\alpha\sigma^{r\phi}](f_1\upsilon^r+f_2\delta\upsilon^r)\nonumber\\
&&-(\frac{1}{\alpha})(n+im\omega)(f_1\upsilon^{\phi}+f_2\delta\upsilon^{\phi})+\frac{1}{\sqrt{\gamma}} B_{\theta}\delta J_r
\end{eqnarray}
and
\begin{eqnarray}\label{v10}
\gamma^{rr}\delta p_{,r}&=&-f_{1,r}(\upsilon^r\upsilon^r)-f_1(\upsilon^r\upsilon^r)_{,r}-2f_{2,r}(\upsilon^r\delta\upsilon^r)-2f_2(\upsilon^r\delta\upsilon^r)_{,r}\nonumber\\
&&-i m f_1(\upsilon^r\upsilon^{\phi})-i m f_2(\upsilon^r\delta\upsilon^{\phi}+\upsilon^{\phi}\delta\upsilon^r)\nonumber\\
&&-(\frac{2}{r}+\Gamma_{rr}^r)\upsilon^r (f_1\upsilon^r+2f_2\delta\upsilon^r)\nonumber\\
&&-\Gamma_{\phi\phi}^r\upsilon^{\phi}(f_1\upsilon^{\phi}+2f_2\delta\upsilon^{\phi})\nonumber\\
&&-(\frac{1}{\alpha})[\Gamma_{t \phi}^r+\omega\Gamma_{\phi\phi}^r+\alpha\gamma_{\phi\phi}\sigma^{r\phi}](f_1\upsilon^{\phi}+f_2\delta\upsilon^{\phi})\nonumber\\
&&-a^r(\delta p+\delta\varepsilon)-(\frac{1}{\alpha})(n+i m\omega)(f_1\upsilon^r+f_2\delta\upsilon^r)\nonumber\\
&&-\frac{1}{\sqrt{\gamma}}[B_{\theta} \delta J_{\phi}+J_{\phi}\delta B_{\theta}]
\end{eqnarray}
Eqs.~(\ref{mhd2a}) \& (\ref{per1a}) yield (see Appendix~B)
\begin{eqnarray}\label{v10b}
\frac{(n+i m \omega)\delta\varepsilon}{\alpha}&=&-2\gamma_{rr} a^r[f_1\upsilon^r+f_2\delta\upsilon^r]-\tilde{\upsilon}\cdot\tilde{\nabla}f_1-\delta\tilde{\upsilon}\cdot\tilde{\nabla}f_2\nonumber\\
&&-2\sigma_{r\phi}[f_1\upsilon^r\upsilon^{\phi}+f_2(\upsilon^r\delta\upsilon^{\phi}+\upsilon^{\phi}\delta\upsilon^r)]\nonumber\\
&&-\gamma_{ij}\delta J^i E^j-\gamma_{\phi\phi} J^{\phi}\delta E^{\phi}
\end{eqnarray}

Notice that because of eqs.~(\ref{vp2}), (\ref{v4}) and (\ref{v7})
\begin{eqnarray}\label{va}
\tilde{\nabla}\cdot \delta \tilde{S}&=&\tilde{\nabla}\cdot(f_1 \tilde{\upsilon})+\tilde{\nabla}\cdot (f_2\delta \tilde{\upsilon})\nonumber\\
&&=\tilde{\upsilon}\cdot\tilde{\nabla}f_1+\delta\tilde{\upsilon}\cdot \tilde{\nabla}f_2
\end{eqnarray}
Eqs.~(\ref{v9}), (\ref{v10}) \& (\ref{v10b}) simplify considerably:
\begin{eqnarray}\label{q9}
\delta p_{,r}&=&-(\frac{\gamma_{rr}}{\alpha})(n+i m\omega+i m\alpha \upsilon^{\phi})f_2\delta\upsilon^r-\gamma_{rr}\Gamma_{\phi\phi}^r\upsilon^{\phi}f_2\delta\upsilon^{\phi}\nonumber\\
&&-(\frac{\gamma_{rr}}{\alpha})[\Gamma_{t \phi}^r+(\omega+\alpha\upsilon^{\phi})\Gamma_{\phi\phi}^r+\alpha\gamma_{\phi\phi}\sigma^{r\phi}](f_1\upsilon^{\phi}+f_2\delta\upsilon^{\phi})\nonumber\\
&&- \gamma_{rr} a^r\{c_s^2\delta\rho-\frac{\alpha(n-im\omega)}{n^2+m^2\omega^2}\nonumber\\
&&\times[\frac{\rho_{,r}\delta\upsilon^r}{1-\tilde{\upsilon}^2}(1-c_s^2+2\tilde{\upsilon}^2c_s^2)+f_2\frac{(\tilde{\upsilon}^2)_{,r}}{1-\tilde{\upsilon}^2}\delta\upsilon^r
\nonumber\\
&&-2(\frac{f_2}{\alpha})\upsilon^{\phi}\delta\upsilon^r\gamma_{rr}(\Gamma_{t\phi}^r+(\omega+\alpha \upsilon^{\phi})\Gamma_{\phi\phi}^r )+i m\upsilon^{\phi}f_{1}\nonumber\\
&&+\gamma_{\phi\phi} J^{\phi}\delta E^{\phi}-\frac{2}{\sqrt{\gamma}}\gamma_{rr}\delta\upsilon^r B_{\theta}J_{\phi}]\}\nonumber\\
&&-\frac{\gamma_{rr}}{\sqrt{\gamma}}(B_{\theta} \delta J_{\phi}+J_{\phi}\delta B_{\theta})\ ,\\
\label{q10}
\delta p_{,\phi}&=&-(\frac{\gamma_{\phi\phi}}{\alpha})(n+im\omega+i m\alpha\upsilon^{\phi})(f_1\upsilon^{\phi}+f_2\delta\upsilon^{\phi})\nonumber\\
&&-\gamma_{\phi\phi}f_{2,r}\upsilon^{\phi}\delta\upsilon^r-\gamma_{\phi\phi}f_2(\upsilon^{\phi}\delta\upsilon^r)_{,r}-\gamma_{\phi\phi}(i m\upsilon^{\phi})(f_2\delta\upsilon^{\phi})\nonumber\\
&&-(\frac{\gamma_{\phi\phi}}{\alpha})[(\frac{2}{r}+\Gamma_{r\phi}^{\phi})\alpha\upsilon^{\phi}+\Gamma_{t r}^{\phi}+\omega\Gamma_{\phi r}^{\phi}\nonumber\\
&&-\omega a_r+\gamma_{rr}\alpha\sigma^{r\phi}]f_2\delta\upsilon^r+\frac{\gamma_{\phi\phi}}{\sqrt{\gamma}} B_{\theta}\delta J_r
\end{eqnarray}
where
\begin{eqnarray}\label{q6}
&&\tilde{\upsilon}^2=\upsilon_i\upsilon^i,\nonumber\\
&&f_{1,\phi}=i m f_1,~~f_{2,r}=(1+c_s^2)\left(\frac{\rho_{,r}}{1-\tilde{\upsilon}^2}+\frac{\rho(\tilde{\upsilon}^2)_{,r}}{(1-\tilde{\upsilon}^2)^2}\right)\nonumber\\
&&f_1\upsilon^{\phi}+f_2\delta\upsilon^{\phi}=\frac{1+c_s^2}{1-\tilde{\upsilon}^2}\upsilon^{\phi}\delta\rho+\frac{1+\tilde{\upsilon}^2}{1-\tilde{\upsilon}^2}f_2\delta\upsilon^{\phi}
\end{eqnarray}
We define here the angular velocity of the flow as
\begin{equation}
\Omega\equiv\omega+\alpha\upsilon^{\phi}
\end{equation}
From eq.~(\ref{vv5}) and (\ref{v4b}), we obtain (see Appendix~A)
\begin{eqnarray}\label{q6t}
&&(n+im\Omega)\delta\rho=-[\frac{\alpha}{r^2}\rho_{,r}+\frac{\rho\Gamma^2\upsilon_{\phi}}{2r^2}G_6(r)](r^2\delta\upsilon^r)\nonumber\\
&&-\frac{i}{m}(n+im\omega+2im\alpha \upsilon^{\phi})[\frac{\rho \Gamma^2\upsilon_{\phi}}{2r^2}](r^2\delta\upsilon^r)_{,r}
\end{eqnarray}

Furthermore, eq.~(\ref{q10}) with the aid of eqs.~(\ref{q6}), (\ref{q6t}) and (\ref{per1a}) becomes
\begin{equation}\label{fin3}
\delta p=N_1+i N_2
\end{equation}
where $N_1$ and $N_2$ are complex expressions that can be found in the Appendix~C. Eq.~(\ref{q9}) with the aid of eq.~ (\ref{q6}) becomes
\begin{equation}\label{fin7}
\delta p_{,r}=\Lambda_1+i\Lambda_2
\end{equation}
where again $\Lambda_1$ and $\Lambda_2$ can be found in the Appendix~ C.

Taking the $r$-derivative of eq.~(\ref{fin3}) and comparing it with eq.~(\ref{fin7}), we obtain two independent equations in the complex plane, namely
\begin{equation}\label{fin7d}
N_{1,r}=\Lambda_1\ \mbox{and}\
N_{2,r}=\Lambda_2\ .
\end{equation}
We will focus on the first equation of (\ref{fin7d}) which, after tedious but straightforward calculations, reduces to the equation
\begin{eqnarray}\label{fin11}
&&[\frac{A}{r^4}(\frac{1+\tilde{\upsilon}^2}{1-\tilde{\upsilon}^2}f_2+\frac{B^2}{4\pi})w_{,r}]_{,r}-\frac{a_rA}{r^4}(\frac{1+\tilde{\upsilon}^2}{1-\tilde{\upsilon}^2}f_2+\frac{B^2}{4\pi})w_{,r}\nonumber\\
&&+a_r(\frac{A}{2r^4})\frac{\rho\tilde{\upsilon}^2(1+c_s^2)}{1-\tilde{\upsilon}^2}w_{,r}-(\frac{A}{2r^4})\frac{\tilde{\upsilon}^2(1+c_s^2)}{1-\tilde{\upsilon}^2}\rho_{,r}w_{,r}\nonumber\\
&&-\rho[(\frac{A}{2r^4})\frac{\tilde{\upsilon}^2(1+c_s^2)}{1-\tilde{\upsilon}^2}w_{,r}]_{,r}\nonumber\\
&&=\frac{m^2}{\Delta}(f_2+\frac{B^2}{4\pi})w\nonumber\\
&&-\frac{G_2(r)}{\Delta}[\frac{m^2}{n^2+m^2\Omega^2}][\frac{1+c_2^2}{1-\tilde{\upsilon}^2}][(\rho_{,r}\alpha+\frac{\rho\upsilon^{\phi}G_6(r)}{2(1-\tilde{\upsilon}^2)})w-\frac{\alpha\rho\tilde{\upsilon}^2}{2(1-\tilde{\upsilon}^2)}w_{,r}]\nonumber\\
&&-(\frac{a^r}{\Delta})\frac{m^2\alpha c_s^2}{n^2+m^2\Omega^2}[(\rho_{,r}\alpha+\frac{\rho\upsilon^{\phi}G_6(r)}{2(1-\tilde{\upsilon}^2)})w-\frac{\alpha\rho\tilde{\upsilon}^2}{2(1-\tilde{\upsilon}^2)}w_{,r}]\nonumber\\
&&-\frac{m^2}{n^2+m^2\omega^2}\frac{\alpha^2\gamma_{rr}a^r}{r^2}\{\frac{1-c_s^2+2c_s^2\tilde{\upsilon}^2}{1-\tilde{\upsilon}^2}\rho_{,r} w\nonumber\\
&&-\frac{2 r^2}{\Delta}\frac{f_2}{\alpha}\upsilon^{\phi}G_3(r)w+f_2\frac{(\tilde{\upsilon}^2)_{,r}}{1-\tilde{\upsilon}^2}w\nonumber\\
&&-[\frac{1+c_s^2}{1-\tilde{\upsilon}^2}\frac{1}{n^2+m^2\Omega^2}][m^2\Omega(\rho_{,r}\alpha+\frac{\rho\upsilon^{\phi}G_6(r)}{2(1-\tilde{\upsilon}^2)})w\nonumber\\
&&-\frac{\rho\tilde{\upsilon}^2}{2(1-\tilde{\upsilon}^2)}w_{,r}(n^2+m^2\Omega\omega+2m^2\Omega\alpha\upsilon^{\phi})]\nonumber\\
&&-\frac{2\tilde{\upsilon}^2}{1-\tilde{\upsilon}^2}f_2 w_{,r}-\frac{3}{4\pi }[B_{\theta} B_{,r}^{\theta}+(\frac{2}{r}+a^r)B^2]w\}\nonumber\\
&&+\frac{\alpha^2a^r}{\Delta}[\frac{1+c_2^2}{1-\tilde{\upsilon}^2}]\frac{m^4\omega}{(n^2+m^2\omega^2)(n^2+m^2\Omega^2)}[(\rho_{,r}\alpha\upsilon^{\phi}+\frac{\rho\tilde{\upsilon}^2 G_6(r)}{2(1-\tilde{\upsilon}^2)})w\nonumber\\
&&-\frac{\alpha\upsilon^{\phi}\rho\tilde{\upsilon}^2}{2(1-\tilde{\upsilon}^2)}w_{,r}]\nonumber\\
&&-\frac{m^2}{4\pi}\frac{B_{\theta} B_{,r}^{\theta}}{n^2+m^2\Omega^2}\{\frac{\alpha\Omega}{r^2}[-\frac{a^r A\upsilon^{\phi}}{r^2}+(\frac{A\upsilon^{\phi}}{r^2})_{,r}]\nonumber\\
&&+\frac{2\alpha^2 a^r}{r^2}+\frac{\alpha\upsilon^{\phi}}{\Delta}G_5(r)\}w\nonumber\\
\end{eqnarray}
where $w\equiv r^2\delta\upsilon^r$ and $G_2, G_3, G_5, G_6$ are given in Appendix~C. Eq.~(\ref{fin11}) is an ordinary second order homogeneous differential equation whose solution yields a criterion for the development of the magnetic RT instability. We verified, that the second equation of (\ref{fin7d}) does not contribute anything more to our study.\\

\section{Stability analysis}

As we discussed in the Introduction, we will now assume that a discontinuous interface develops at the position of the ISCO
\begin{equation}\label{z1a}
r_{\rm ISCO}=M\{3+Z_2-[(3-Z_1)(3+Z_1+2Z_2)]^{1/2}\}\end{equation}
where,
\begin{eqnarray}
Z_1&\equiv&1+(1-\frac{a^2}{M^2})^{1/3}[(1+\frac{a}{M})^{1/3}+(1-\frac{a}{M})^{1/3}],~~\mbox{~and~}\nonumber\\
Z_2&\equiv&(3\frac{a^2}{M^2}+Z_1^2)^{1/2}\nonumber
\end{eqnarray}
Outside $r_{\rm ISCO}$ there is an equatorial disk of plasma with circular Keplerian velocity with respect to ZAMOs equal to
\begin{equation}\label{bar1}
\upsilon^{\phi}=\sqrt{M}\frac{(r^2-2a\sqrt{r M}+a^2)}{r\Delta^{1/2}(r^{3/2}+a \sqrt{M})}
\end{equation}
\citep[][for prograde disk rotation]{a5}.

We will assume that our physical quantities $\rho$, $p$, and $B$ are constant inside and outside $r_{\rm ISCO}$ (at least in its vicinity), but change discontinuously across $r_{\rm ISCO}$.  The radial velocity perturbations $\delta\upsilon^r$ and the total pressure $(p+B^2/8\pi)$ are continuous across the interface between the two fluids, but $\rho$, $B^2$ and $(\delta\upsilon^r)_{,r}$ in general are not.

We make here a further simplifying assumption that the interface lies far from the black hole horizon, namely that $M/r_{\rm ISCO}\ll 1$ (this is obviously not valid for fast rotating black holes where the ISCO approaches the horizon). Under that approximation, we expand in powers of $M/r$ and keep terms only up to $1/r^3$.  Eq.~(\ref{fin11}) now takes the form
\begin{eqnarray}\label{dif4}
&&\frac{d^2 w}{dr^2}+\{-\frac{M[7(1+c_s^2)+2u_A^2]}{2r^2(1+c_s^2+u_A^2)}\nonumber\\
&&+\frac{5M^2(1+c_s^2)[7(1+c_s^2)+2u_A^2]}{4r^3(1+c_s^2+u_A^2)^2} \}\frac{d w}{dr}\nonumber\\
&&+\{ -\frac{m^2}{r^2}-\frac{m^2M(1+c_s^2+4u_A^2)}{2r^3(1+c_s^2+u_A^2)}\} w=0
\end{eqnarray}
inside and outside the discontinuous interface, where $u_A^2\equiv B^2/(4\pi\rho)$.
Eq.~(\ref{dif4}) admits two independent solutions $w_{1}(r), w_{2}(r)$ that apply inside and outside the interface respectively (see Appendix~D). Notice that we haven't made here the assumption of negligibly small magnetic field as we did in  Paper~I.

For any physical quantity $f$ discontinuous across $r_{\rm ISCO}$, we now define
\begin{equation}\label{fin11a}
{\cal D} \{f\}\equiv f_{(2)}-f_{(1)}\ ,\
{\cal P} \{f\}\equiv f_{(2)}+f_{(1)}\ ,
\end{equation}
where $f_{(1)}\equiv f(r_{\rm ISCO}-\epsilon)$ and $f_{(2)}\equiv f(r_{\rm ISCO}+\epsilon)$. The two independent solutions $w_{1}(r), w_{2}(r)$ are then inserted in the full eq.~(\ref{fin11}) which, at the discontinuous interface yields
\begin{eqnarray}\label{vfin11c}
&&\frac{A}{r^4}{\cal D}\{(\frac{1+\tilde{\upsilon}^2}{1-\tilde{\upsilon}^2}f_2+\frac{B^2}{4\pi})w_{,r}\}-(\frac{A}{2r^4})\frac{(1+c_s^2)\tilde{\upsilon}^2}{1-\tilde{\upsilon}^2}{\cal D}\{\rho w_{,r}\}\nonumber\\
&&-\rho{\cal D}\{(\frac{A}{2r^4})\frac{(1+c_s^2)\tilde{\upsilon}^2}{1-\tilde{\upsilon}^2} w_{,r}\}\nonumber\\
&=&-\frac{\alpha G_2(r)}{\Delta}[\frac{m^2(\alpha\upsilon^{\phi})}{n^2+m^2\Omega^2}][\frac{1+c_2^2}{1-\tilde{\upsilon}^2}]{\cal D}\{\rho w\}\nonumber\\
&&-\frac{\alpha a^r}{\Delta}\frac{m^2\alpha c_s^2}{n^2+m^2\Omega^2}{\cal D}\{\rho w\}\nonumber\\
&&-\frac{m^2}{n^2+m^2\omega^2}\frac{\alpha^2a^r}{\Delta}\nonumber\\
&&\times\ \{\frac{1-c_s^2+2\tilde{\upsilon}^2 c_s^2}{1-\tilde{\upsilon}^2} {\cal D}\{\rho w\}
-\frac{3}{8\pi }{\cal D}\{B^2 w\}\nonumber\\
&&-[\frac{1+c_s^2}{1-\tilde{\upsilon}^2}]\frac{m^2\Omega(\alpha\upsilon^{\phi})}{n^2+m^2\Omega^2}{\cal D}\{\rho w\}\}\nonumber\\
&&+\frac{\alpha^2 a^r}{\Delta}[\frac{1+c_s^2}{1-\tilde{\upsilon}^2}]\frac{m^4\omega(\alpha\upsilon^{\phi})}{(n^2+m^2\omega^2)(n^2+m^2\Omega^2)}{\cal D}\{\rho w\}\nonumber\\
&&-\frac{m^2 r^2}{8\pi(n^2+m^2\Omega^2)}{\cal D}\{B^2 w\}\cdot\nonumber\\
&&\{\frac{2\alpha^2 a^r}{r^2}+\frac{\alpha\Omega}{r^2}[-\frac{a^rA\upsilon^{\phi}}{r^2}+(\frac{A\upsilon^{\phi}}{r^2})_{,r}]+\frac{\alpha \upsilon^{\phi}G_5(r)}{\Delta}\}\nonumber\\
\end{eqnarray}
Inserting eqs.~(\ref{dif5}) in eq.~(\ref{vfin11c}), we end up with an equation of the form
\begin{equation}\label{master1}
R=-\frac{m^2}{n^2+m^2\omega^2}[L_1+\frac{m^2\tilde{L_1}}{n^2+m^2\Omega^2}]-\frac{m^2}{n^2+m^2\Omega^2}[L_2+\tilde{L}_2]
\end{equation}
where the expressions for $R$, $L_1$, $\tilde{L}_1$,  $L_2$ and $\tilde{L}_2$ can be found in Appendix~E. Next, we write eq.~(\ref{master1}) as
\begin{eqnarray}\label{RT1a}
&&n^4+n^2m^2[\omega^2+\Omega^2+\frac{L_1+L_2}{R}+\frac{\tilde{L}_2}{R}]\nonumber\\
&&+m^4[\omega^2\Omega^2+\frac{\Omega^2 L_1+\omega^2 L_2}{R}+\frac{\tilde{L}_1}{R}+\frac{\omega^2 \tilde{L}_2}{R}]=0
\end{eqnarray}
This is the main equation of our analysis. Its roots
\begin{eqnarray}\label{RT2}
n_{1,2}^2&=&-\frac{m^2}{2}\left[\omega^2+\Omega^2+\frac{L_1+L_2}{R}+\frac{\tilde{L}_2}{R}\right.\nonumber\\
&& \mp\left([\omega^2+\Omega^2+\frac{L_1+L_2}{R}+\frac{\tilde{L}_2}{R}]^2\right. \nonumber\\
&&\left.\left. -4[\omega^2\Omega^2+\frac{\Omega^2 L_1+\omega^2 L_2}{R}+\frac{1}{ R}(\tilde{L}_1+\omega^2\tilde{L}_2)]\right)^{1/2}\right]\nonumber\\
\end{eqnarray}
characterize the time evolution of our perturbations according to eq.~(\ref{per1a}). Whenever either one of $n_1^2, n_2^2$ is found to be positive, the system will be unstable to the development of the magnetic RT instability. Stability requires that both roots are negative.

At this point we would like to notice that, in our present work, we expanded eq.~(\ref{fin11}) up to third order in terms of  $M/r$, whereas previously, in our Paper~ I, we expanded in terms of $a/M$. In the limit of no black hole rotation, the two approaches yield small differences in the denominator $R$ of eq.~(\ref{RT2}).

Let us here study the stability of a simple configuration at $r=r_{\rm ISCO}$. As we have already said, the density drops inside the ISCO \citep[e.g.][]{Petal10} because of a corresponding increase in the accretion velocity, from a value $v_{r\ {\rm ISCO}}$ just outside the ISCO to $v_{{\rm f\!f}\ {\rm ISCO}}$ just inside the ISCO .
Thus, mass conservation across the ISCO requires that
\begin{eqnarray}\label{art1}
&&\rho_{(1)}=\rho,~~\rho_{(2)}= \rho \left(\frac{v_{{\rm f\!f}\ {\rm ISCO}}}{v_{r\ {\rm ISCO}}}\right)>\rho .
\end{eqnarray}
Let us next assume that there is a significant uniform vertical magnetic field $B$ accumulated inside the ISCO, and no magnetic field outside, i.e.
\begin{eqnarray}\label{art2}
&&B^{\theta}_{(1)}\equiv B\ ,~~B^{\theta}_{(2)}\approx 0
\end{eqnarray}
Pressure balance across the interface requires that $\rho_{(1)} c_s^2+B^2/(8\pi)=\rho_{(2)} c_s^2$, and therefore
\begin{eqnarray}\label{art3}
&&c_s^2=\frac{p}{\rho}=\frac{B^2}{8\pi\rho}\left[\frac{v_{{\rm f\!f}\ {\rm ISCO}}}{v_{r\ {\rm ISCO}}}-1\right]^{-1}
\ .
\end{eqnarray}

We can now calculate the stability regime in which both roots of eq.~(\ref{RT1a}) are negative. Since we are interested in the global flow disruption by the RT instability, we first consider the mode $m=1$ (higher $m$ modes are also considered in Table~1). We also take $v_{{\rm f\!f}\ {\rm ISCO}}=3v_{r\ {\rm ISCO}}$. The second root is found to be negative for all magnetic field values. The first root is negative only for small values of the magnetic field, and becomes positive at some finite value of $B$. In Fig.~1 we plot with a continuous red line the maximum value of $B/\sqrt{4\pi\rho}$ that is stable to the development of the magnetic RT instability as a function of the dimensionless black hole spin parameter $a/M$. We observe that, similarly to the case of no disk rotation studied in Paper~I, a non-rotating non-accreting (Schwarzschild) black hole cannot stably hold any finite amount of magnetic field. As we noted, our present stability analysis differs slightly from that in Paper~I. For black hole spins beyond about $a/M\gsim 0.8$, the ISCO approaches the black hole horizon, and the approximation used to obtain the red line (namely that $M/r_{\rm ISCO}\ll 1$) breaks down.

\section{Astrophysical implications}

\begin{figure}
 \centering
 \includegraphics[width=8cm,height=8cm,angle=0.0]{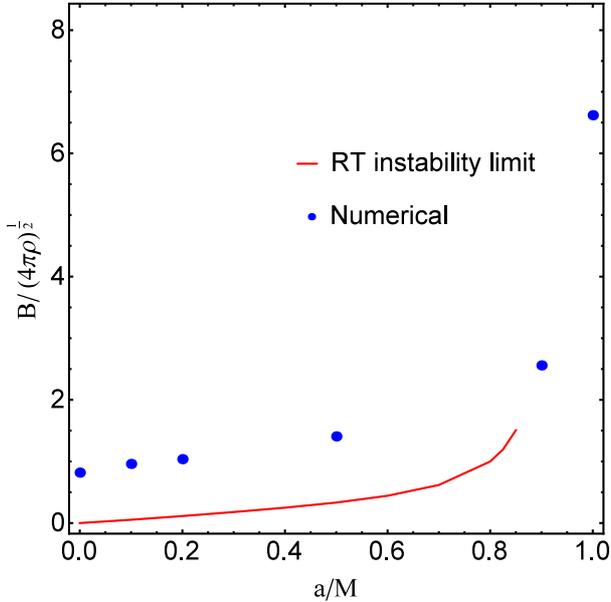}
\caption{Limiting values of the accumulated dimensionless magnetic field $B/\sqrt{4\pi \rho}$ as a function of the dimensionless black hole spin $a/M$. Red line: RT instability limit for the case of no accretion.  Blue points: Numerical simulations \citep[adapted from][via eq.~\ref{Bpsi}]{TMN12}. Notice that the RT instability growth timescales at the blue points are one order of magnitude longer than the corresponding free-fall timescales from the ISCO onto the black hole horizon.}
\label{spectrum profilef3}
\end{figure}

The idealized conditions of a Keplerian disk with a discontinuous interface at the ISCO are rather different from the actual conditions in a real astrophysical accretion flow. If the plasma is ideal, magnetic flux is carried along by the flow in such a way that it conserves the flux to mass ratio $B/\sigma$, where $\sigma$ is the surface density in the disk. As accretion proceeds through the ISCO and the density drops, $B$ also drops, thus the conditions discussed in the previous section do not develop at the ISCO.

Such conditions (namely a drop in density with an increase in the accumulated magnetic field) develop instead on the black hole horizon where our formalism does not apply ($M/r$ is of order unity). We will, thus, adopt a discontinuous configuration at the ISCO and compare our conclusions with the results of GR MHD numerical simulations. As an example, figure~4a of \cite{TMN12} shows the maximum dimensionless magnetic flux $\phi_{\rm BH}\equiv \Phi_{\rm BH}/\dot{M}^{1/2}r_{\rm g}$
accumulated on the black hole horizon as a function of the dimensionless black hole spin (for prograde flow rotation). In their notation, $\Phi_{\rm BH}$ is the actual accumulated magnetic flux, $\dot{M}=2\pi \rho rh v_r$ is the mass accretion rate, $h$ is the disk thickness, and $v_r$ is the accretion velocity. In order to connect $\phi_{\rm BH}$ to $B/\sqrt{4\pi\rho}$ at the ISCO, we set $\Phi_{\rm BH}=\Phi_{\rm ISCO}\equiv \pi r_{\rm ISCO}^2 B$. We also calculate $\dot{M}$ at the ISCO. Thus we obtain
\begin{equation}
\frac{B}{\sqrt{4\pi\rho}}
\approx \frac{\phi_{\rm BH}}{\pi}\left(\frac{r_{\rm ISCO}}{r_{\rm g}}\right)^{-5/4}\left(\frac{h}{2r}\right)^{1/2}\ .
\label{Bpsi}
\end{equation}
The ratio $h/r$ around the inner edge of the disk is not known. We can roughly estimate it from figure~3a of \cite{TMN12} as $h/2r\sim 0.2$. The simulation points resulting from eq.~(\ref{Bpsi}) (blue points in Fig.~1) lie above the RT instability limit (red line) for black hole spins below about $0.9M$, thus, according to our present analysis, they must be unstable to the development of the RT instability.

It is important to realize that the instability does not manifest itself instantaneously, but grows with an exponential e-folding time $t_{\rm inst}\equiv 1/n_1$, where $n_1^2$ is the positive first root of eq.~(\ref{RT2}) in that region. During that time, accretion proceeds and brings the accumulated magnetic flux towards the horizon on a free-fall timescale of the order of $t_{\rm ff}\equiv \sqrt{2}(r_{\rm ISCO}^{3/2}-r_{\rm bh}^{3/2})/(3\sqrt{GM})$, where $r_{\rm bh}$ is the radius of the black hole horizon (see Table~1; the timescale calculation is Newtonian). For the magnetic field parameters that correspond to the blue (simulation) points above the red line we found that the instability growth timescales are close to the (classical) free-fall accretion times from the ISCO onto the black hole horizon. This implies that the RT instability has enough time to begin manifesting itself.

\begin{table}
\caption{Instability growth timescales $t_{\rm inst}$ (in italics) for various values of $m$ for the simulation points at $a=0, 0.1M, 0.2M, 0.5M$ above the red line. Also shown the corresponding classical free-fall times $t_{\rm ff}$. Times in units of $GM/c^3$.}
\centering
\begin{tabular}{cccccccc}
\toprule
$a$	& $t_{\rm ff}$ & $m=1$ & $2$ & $3$ & $5$ & $10$ & $20$ \\
\midrule
$0$	      & 5.6 & {\it 5.1}			&  {\it 3.7} &  {\it 3.3} &   {\it 3.6} &  {\it 3.8} &  {\it 4.2}    \\
$0.1M$	& 5.1 &  {\it 4.7}			&  {\it 3.3} &  {\it 2.9} &  {\it 2.6}  &  {\it 3.2}  &   {\it 3.8}    \\
$0.2M$	& 4.4 &  {\it 4.5}			&  {\it 3.2} &  {\it 2.8} &   {\it 2.6} &  {\it 3.3}  &   {\it 3.7}    \\
$0.5M$	& 2.9 &  {\it 4.3}			&  {\it 3.1} &  {\it 2.8} &  {\it 2.9}  &   {\it 3.7} &  {\it 4.6}     \\
\bottomrule
\end{tabular}
\end{table}

The latter result is rather interesting. \cite{TMN12} conclude that, in steady state, the black hole is saturated with magnetic flux, and the magnetic field is so strong that it obstructs the accretion and leads to a magnetically-arrested disk (MAD). It is very interesting that the maximum accumulated dimensionless magnetic flux is found to be roughly equal to its equipartition value independent of the black hole spin (for prograde flow rotation). Our present results suggest that the accumulation of the magnetic field may also be limited by the RT instability. In other words, the process of accretion and magnetic flux accumulation on the black hole horizon is probably disrupted both by the strong magnetic field and the RT instability.

In summary, our investigation of the magnetic Rayleigh-Taylor instability in this series of two papers showed that the amount of magnetic flux that can be stably accumulated inside the ISCO of a Keplerian accretion disk around a black hole is small for a slowly spinning black hole, and increases for higher black hole spins. We also found that, for black hole spins $a<0.9M$ for which our present analysis is valid, the disk reaches a magnetically arrested state (MAD) and the accretion flow is disrupted at about the same time that the magnetic flux accumulation onto the black hole horizon is disrupted by the Rayleigh-Taylor instability.

\section*{Acknowledgements}

We thank the anonymous referee who pointed out several mistakes in the original version of the paper that led us to repeat our calculations.


\bibliographystyle{mn2e}

\section*{Appendix~A}
We derive here the law of conservation of mass (or baryon number)  in 3+1 formalism. According to our notation,
\begin{eqnarray}\label{p1}
&& u^\mu \equiv \frac{{\rm d}x^\mu_{\rm fluid}}{{\rm d}\tau_{\rm fluid}}\ \ \mbox{is the fluid 4-velocity}\nonumber\\
&& U^\mu \equiv \frac{{\rm d}x^\mu_{\rm ZAMO}}{{\rm d}\tau_{\rm ZAMO}}\ \ \mbox{is the ZAMO 4-velocity}\nonumber\\
&& \tilde{v}^i \equiv \frac{{\rm d}x^i_{\rm fluid\ w.r.\ to\ ZAMO}}{{\rm d}\tau_{\rm ZAMO}}\ \ \mbox{is the fluid 3-velocity wr to ZAMO}\nonumber\\
&& \alpha \equiv \frac{{\rm d}\tau_{\rm ZAMO}}{{\rm d}t}\nonumber\\
&& u^0 \equiv \frac{{\rm d}t}{{\rm d}\tau_{\rm fluid}}=
\frac{{\rm d}\tau_{\rm ZAMO}}{{\rm d}\tau_{\rm fluid}}\frac{{\rm d}t}{{\rm d}\tau_{\rm ZAMO}}=
\frac{\Gamma}{\alpha}\ ,\nonumber
\end{eqnarray}
where ${\rm d}\tau_{\rm ZAMO}/{\rm d}\tau_{\rm fluid}=\Gamma$ is the fluid Lorentz factor w.r. to ZAMO observers. Obviously,
\begin{eqnarray}\label{p2}
&& {\rm d}x^i_{\rm fluid}={\rm d}x^i_{\rm ZAMO}+{\rm d}x^i_{\rm fluid\ wr\ to\ ZAMO}
\end{eqnarray}
thus
\begin{eqnarray}\label{p3}
u^i & = & (\tilde{v}^i+U^i)\left(\frac{{\rm d}\tau_{\rm ZAMO}}{{\rm d}\tau_{\rm fluid}}\right)\nonumber\\
& \equiv & (\tilde{v}^i+U^i)\Gamma \ ,
\end{eqnarray}
According to \cite{xn9}, the mass conservation becomes
\begin{eqnarray}\label{p4}
&& (\rho u^{\mu})_{;\mu}=(\rho u^0)_{;0}+(\rho u^i)_{;i}=\nonumber\\
&&(\rho \Gamma/\alpha)_{;t}+
(\rho \Gamma (\tilde{v}^i+U^i))_{;i}=0\ ,
\end{eqnarray}
or equivalently
\begin{eqnarray}\label{d2}
&& [(\rho\Gamma/\alpha)_{,t}+U^i (\alpha(\rho\Gamma/\alpha))_{,i}]+
\rho\Gamma (U^i)_{;i} + (\rho\Gamma \tilde{v}^i)_{;i}=\nonumber\\
&& \alpha D_\tau (\rho\Gamma/\alpha) +\rho\Gamma\theta+\tilde{\nabla}\cdot (\rho\Gamma \tilde{\upsilon})=\nonumber\\
&& D_{\tau}(\rho\Gamma)+\rho\Gamma\theta+\tilde{\nabla}\cdot (\rho\Gamma\tilde{\upsilon})=0
\label{A1}
\end{eqnarray}
(we remind the reader that the expansion $\theta\equiv \tilde{\nabla}\cdot \tilde{U}$ is equal to zero in Kerr space time).
We perturb Eq.~(\ref{A1}) and find
\begin{eqnarray}\label{b1}
&&D_{\tau}(\Gamma \delta\rho+\rho \delta\Gamma)+\tilde{\nabla}\cdot(\rho\Gamma\delta\tilde{\upsilon}+\tilde{\upsilon}\Gamma \delta\rho+\tilde{\upsilon}\rho\delta\Gamma)=0
\end{eqnarray}
or
\begin{eqnarray}\label{b1a}
&&D_{\tau}\delta\rho+\tilde{\nabla}\cdot(\rho\delta\tilde{\upsilon})+\tilde{\nabla}\cdot(\tilde{\upsilon}\delta\rho)\nonumber\\
&&=-\frac{\delta\rho D_{\tau}\Gamma+\rho D_{\tau}\delta \Gamma}{\Gamma}
-\frac{\rho\Gamma^2\tilde{v}\cdot\tilde{\nabla}\delta\tilde{v}^2}{2}
\end{eqnarray}
where $\tilde{\upsilon}^2=\upsilon_i\upsilon^i$, with $\tilde{\upsilon}\rightarrow\upsilon^i=(\upsilon^r,0,\upsilon^\phi)$ and
\begin{equation}\label{b1b}
\tilde{\nabla}\cdot(\rho\delta\tilde{\upsilon})=\frac{1}{\sqrt{\gamma}}\frac{\partial}{\partial x^i}[\sqrt{\gamma}\rho\delta\upsilon^i]
\end{equation}

Taking into account that $\Gamma=(1-\tilde{\upsilon}^2)^{-1/2}$ and that the Fermi derivative includes $t$ and $\phi$ derivatives, we find immediately that $D_{\tau}\Gamma=0$.
Furthermore,
\begin{equation}\label{f5}
\delta \Gamma=\frac{1}{2}(1-\tilde{\upsilon}^2)^{-3/2}\delta \tilde{\upsilon}^2
=\frac{\Gamma^3}{2}(\delta\tilde{\upsilon}^2)=\Gamma^3\tilde{\upsilon}\cdot\delta\tilde{\upsilon}\ ,
\end{equation}
and
\begin{equation}\label{f5x}
D_{\tau}\delta\Gamma=\Gamma^3\tilde{\upsilon}\cdot D_{\tau}\delta\tilde{\upsilon}
\end{equation}

Inserting all values in eq.~(\ref{b1a}), we find
\begin{eqnarray}\label{c2}
&&D_{\tau}\delta \rho+\delta\tilde{\upsilon}\cdot\tilde{\nabla}\rho+\tilde{\upsilon}\cdot\tilde{\nabla}\delta\rho+\delta\rho\tilde{\nabla}\cdot\tilde{\upsilon}+\rho\tilde{\nabla}\cdot\delta\tilde{\upsilon}\nonumber\\
&&=-\frac{\rho\Gamma^2}{2}\tilde{\upsilon}\cdot D_{\tau}\delta\tilde{\upsilon}-\frac{\Gamma^2}{2}\rho\tilde{\upsilon}\cdot\tilde{\nabla}\tilde{\delta \upsilon}^2
\end{eqnarray}
Notice that since $\tilde\upsilon$ has only a $\phi$-component, and $\tilde{\upsilon}^2$ does not depend on $\phi$, $\tilde{\upsilon}\cdot \tilde{\nabla}\tilde{\upsilon}^2=0$. We have also assumed for simplicity that $\tilde{\nabla}\cdot \tilde{\upsilon}=0$ (eq.~\ref{v4}). Thus, eq.~(\ref{c2}) becomes
\begin{equation}\label{c2a}
D_{\tau}\delta \rho+\tilde{\upsilon}\cdot\tilde{\nabla}\delta\rho
=-\delta\tilde{\upsilon}\cdot\tilde{\nabla}\rho-\frac{\rho\Gamma^2}{2}\tilde{\upsilon}\cdot D_{\tau}\delta\tilde{\upsilon}-\frac{\Gamma^2}{2}\rho\tilde{\upsilon}\cdot\tilde{\nabla}\tilde{\delta \upsilon}^2
\end{equation}

For our further calculations we use mathematical formulas from \cite{uz05}, \cite{JP03} and compute
\begin{eqnarray}\label{b2a}
&&[\delta\tilde{\upsilon}^2]_{,\phi}=2\gamma_{\phi\phi}\upsilon^{\phi}\delta\upsilon_{,\phi}^{\phi}=2im\upsilon_{\phi}\delta\upsilon^{\phi}\ ,\nonumber\\
&&\label{b2b}
\tilde{\upsilon}\cdot\tilde{\nabla}(\delta\tilde{\upsilon}^2)=
\upsilon^\phi\partial_\phi(\delta\tilde{\upsilon}^2)=\upsilon^\phi[2im\upsilon_{\phi}\delta\upsilon^{\phi}]\ .
\end{eqnarray}
In eq.~(\ref{c2a}), the term $\tilde{\upsilon}\cdot D_{\tau}\delta\tilde{\upsilon}$ becomes
\begin{eqnarray}\label{c2b}
&&\tilde{\upsilon}\cdot D_{\tau}\delta\tilde{\upsilon}=\upsilon_{\phi}[D_{\tau}\delta \tilde{\upsilon}]^{\phi}\nonumber\\
&&=\upsilon_{\phi}\{[(\delta \upsilon^{\phi})_{,k}+\Gamma_{kl}^{\phi}\delta\upsilon^l] U^k-U^{\phi}(a_r\delta\upsilon^r)\}
\end{eqnarray}
After straightforward calculations, eq.~(\ref{c2b}) with the aid of eq.~(\ref{k3x}) becomes
\begin{equation}\label{c2c}
\tilde{\upsilon}\cdot D_{\tau}\delta\tilde{\upsilon}=\frac{\upsilon_{\phi}}{\alpha}[(n+im\omega)\delta\upsilon^{\phi}+\frac{G_6(r)}{r^2}(r^2\delta\upsilon^r)]
\end{equation}
where $\delta \upsilon^{\phi}$ is given by eq.~(\ref{v4b}) and
\begin{equation}\label{c2d}
G_6(r)\equiv \Gamma_{t r}^{\phi}+\omega\Gamma_{\phi r}^\phi-\omega a_r
\end{equation}
Eventually, from eqs.~(\ref{c2a})-(\ref{c2d}) we have
\begin{eqnarray}\label{c2e}
&&(n+im\Omega)\delta\rho=-[\frac{\alpha}{r^2}\rho_{,r}+\frac{\rho\Gamma^2\upsilon_{\phi}}{2r^2}G_6(r)](r^2\delta\upsilon^r)\nonumber\\
&&-\frac{i}{m}(n+im\omega+2im\alpha \upsilon^{\phi})[\frac{\rho \Gamma^2\upsilon_{\phi}}{2r^2}](r^2\delta\upsilon^r)_{,r}
\end{eqnarray}
which is eq.~(\ref{q6t}) in the text.



\section*{Appendix~B}

We derive here eq.~(\ref{v10b}) in the text. We re-write eq.~(\ref{mhd2a}) as follows:
\begin{equation}\label{s1}
D_{\tau}\delta\varepsilon =-2\delta\tilde{S}\cdot \tilde{a}-\tilde{\nabla}\cdot \delta\tilde{S}-\tilde{\sigma}\cdot\delta \tilde{W}
-\delta\tilde{J}\cdot\tilde{E}-\tilde{J}\cdot\delta\tilde{E}\ ,
\end{equation}
where the Fermi derivative of the scalar $\delta \varepsilon$ is
\begin{eqnarray}\label{s2}
D_{\tau}\delta\varepsilon&=&\frac{1}{\alpha}[\delta \varepsilon_{,\mu}(\alpha U^\mu+\beta^\mu)-\gamma^{ij}\beta_i\delta \varepsilon_{,j}\nonumber\\
&=&\frac{1}{\alpha}[n+im\omega]\delta \varepsilon
\end{eqnarray}
Eq.~(\ref{s2}), is written as
\begin{eqnarray}\label{s3}
\frac{1}{\alpha}[n+im\omega]\delta \varepsilon&=&-2\gamma_{ij}a^i\delta S^j-\tilde{\nabla}\cdot \delta\tilde{S}-\sigma_{ij}\delta \hat{W}^{ij}\nonumber\\
&-&\gamma_{ij}\delta J^i E^J-\gamma_{ij}J^i\delta E^J
\end{eqnarray}
The third of eqs.~(\ref{mhd2b}), with the aid of eqs.~(\ref{v7}), gives
\begin{equation}\label{s4}
\delta \tilde{S} = f_1\tilde{\upsilon}+f_2\delta\tilde{\upsilon}
\end{equation}
where eqs.~(\ref{v7}) reduce to
\begin{eqnarray}\label{s5}
f_1&\equiv&(\delta\rho+\delta p)\Gamma^2+(\rho+p)\delta\Gamma^2\nonumber\\
&=&\frac{(1+c_s^2)}{1-\tilde{\upsilon}^2}\delta \rho+2\frac{(1+c_s^2)}{1-\tilde{\upsilon}^2}u_k\delta\upsilon^k\nonumber\\
f_2&\equiv&(\rho+p)\Gamma^2=\frac{(1+c_s^2)}{1-\tilde{\upsilon}^2}\rho
\end{eqnarray}
for an isothermal fluid with $c_s^2=p/\rho$. Furthermore, under our assumption that the fluid is incompressible (namely $\tilde{\nabla}\cdot\tilde{\upsilon}=0$ and $\tilde{\nabla}\cdot\delta \tilde{\upsilon}=0$), we obtain
\begin{eqnarray}\label{s6}
\tilde{\nabla}\cdot\delta\tilde{S}&=&\tilde{\nabla}\cdot(f_1\tilde{\upsilon})+\tilde{\nabla}\cdot(f_2\delta\tilde{\upsilon})\nonumber\\
&=&f_1\tilde{\nabla}\cdot\tilde{\upsilon}+\tilde{\upsilon}\cdot\tilde{\nabla}f_1+f_2\tilde{\nabla}\cdot\delta \tilde{\upsilon}+\delta \tilde{\upsilon}\cdot\tilde{\nabla}f_2\nonumber\\
&=&\tilde{\upsilon}\cdot\tilde{\nabla}f_1+\delta\tilde{\upsilon}\cdot\tilde{\nabla}f_2
\end{eqnarray}
and thus eq.~(\ref{s3}) reads
\begin{eqnarray}\label{s7}
\frac{1}{\alpha}[n+im\omega]\delta \varepsilon&=&-2\gamma_{ij}a^i\delta S^j-\gamma^{ij}\upsilon_i[\nabla f_1]_j+\gamma^{ij}\delta\upsilon_i [\nabla f_2]_j\nonumber\\
&-&\sigma_{ij}\delta \hat{W}^{ij}-\gamma_{ij}\delta J^i E^J-\gamma_{ij}J^i\delta E^J
\end{eqnarray}
On the equatorial plane, there exist only one non-zero component of the acceleration ($a^r$) and one component of the shear tensor ($\sigma_{r\phi}$). Also, we assume that the current $J^{\mu}$ has only one non-zero component ($J^{\phi}$), and similarly for the velocity $\tilde{\upsilon}^i$ ($\tilde{\upsilon}^{\phi}$). Eq.~(\ref{s7}), now reads:
\begin{eqnarray}\label{s8}
\frac{1}{\alpha}[n+im\omega]\delta \varepsilon&=&-2\gamma_{rr}a^r\delta S^r-\gamma^{\phi\phi}\upsilon_{\phi}[\nabla f_1]_{\phi}+\gamma^{ij}\delta\upsilon_i [\nabla f_2]_j\nonumber\\
&-&2\sigma_{r\phi}\delta \hat{W}^{r\phi}-\gamma_{ij}\delta J^{i} E^{j}-\gamma_{\phi\phi}J^{\phi}\delta E^{\phi}
\end{eqnarray}
where from eq.~(\ref{s4}) with $i=r$ we obtain the $\delta S^r$ and from the forth of eqs.~(\ref{mhd2b}) we obtain
\begin{equation}\label{s9}
\delta\hat{W}^{r\phi}=f_1\upsilon^r\upsilon^{\phi}+f_2(\upsilon^r\delta\upsilon^{\phi}+\upsilon^{\phi}\delta\upsilon^r)
\end{equation}

From eqs.~(\ref{s8}), (\ref{s4}) and (\ref{s9}) we find
\begin{eqnarray}\label{s9a}
\frac{1}{\alpha}(n+i m \omega)\delta\varepsilon&=&-2\gamma_{rr} a^r[f_1\upsilon^r+f_2\delta\upsilon^r]-\tilde{\upsilon}\cdot\tilde{\nabla}f_1-\delta\tilde{\upsilon}\cdot\tilde{\nabla}f_2\nonumber\\
&-&2\sigma_{r\phi}[f_1\upsilon^r\upsilon^{\phi}+f_2(\upsilon^r\delta\upsilon^{\phi}+\upsilon^{\phi}\delta\upsilon^r)]\nonumber\\
&&-\gamma_{ij}\delta J^i E^j-\gamma_{\phi\phi} J^{\phi}\delta E^{\phi}
\end{eqnarray}

Further, we will compute the term $-2\gamma_{rr} (a^r f_2)\delta\upsilon^r$ in eq.~(\ref{s9a}) using the zero-order eq.~(\ref{x1a}).

We consider small perturbations of the form (\ref{per1}) with (\ref{per1a})
In this case we have
\begin{eqnarray}\label{s12}
&&(\upsilon_k\delta\upsilon^k)_{,\phi}=\upsilon_k\delta\upsilon_{,\phi}^k=im \upsilon^k\delta\upsilon^k\nonumber\\
&&f_{1,\phi}\equiv\frac{\partial f_1}{\partial\phi}=im f_1\nonumber\\
&&f_{2,\phi}=0\nonumber\\
&&\delta\upsilon_{,\phi}^i=im\delta\upsilon^i
\end{eqnarray}
and
\begin{equation}\label{s13}
\tilde{\nabla}\cdot\delta\tilde{\upsilon}=0\Rightarrow -im \delta\upsilon^{\phi}=\frac{1}{r^2}(r^2\delta\upsilon^r)_{,r}\equiv\chi
\end{equation}

For our further computations we will use some of the zero-order eqs.~(\ref{x1a}) in the text. From the zeroth order MHD equations we keep only
\begin{equation}\label{x1}
D_{\tau}\tilde{S}+\tilde{\sigma}\cdot \tilde{S}+\varepsilon\tilde{a}+\tilde{\nabla}\cdot \tilde{W}+\tilde{W}\cdot\tilde{a}=(\rho_e\tilde{E}+\tilde{J}\times \tilde{B})
\end{equation}
where
\begin{eqnarray}\label{x2}
&&\tilde{S}=f_2\tilde{\upsilon},~~\tilde{W}=f_2\tilde{\upsilon}\otimes\tilde{\upsilon}+p\tilde{\gamma}\nonumber\\
&&\varepsilon=\Gamma^2(\rho+p\tilde{\upsilon}^2)=\Gamma^2\rho(1+c_s^2\tilde{\upsilon}^2)
\end{eqnarray}

\begin{eqnarray}\label{x3}
&&D_{\tau} S^{\beta}=(S_{,\mu}^{\beta}+\Gamma_{\mu\nu}^{\beta} S^{\nu})U^{\mu}-U^{\beta}(a_k S^k)\nonumber\\
&&\tilde{\nabla}\cdot\tilde{S}\equiv\tilde{\nabla}\cdot(f_2\tilde{\upsilon})=\frac{1}{\sqrt{\gamma}}\frac{\partial}{\partial x^i}[\sqrt{\gamma} f_2\upsilon^i]\nonumber\\
&&\tilde{\nabla}\cdot\tilde{W}\equiv\tilde{\nabla}\cdot[f_2\tilde{\upsilon}\otimes \tilde{\upsilon}]+\tilde{\nabla}\cdot [p\tilde{\gamma}]\nonumber\\
&&=\frac{1}{\sqrt{\gamma}}\frac{\partial}{\partial x^i}[\sqrt{\gamma} f_2\upsilon^i\upsilon^j]+\Gamma_{kl}^j(f_2\upsilon^k\upsilon^l)+\gamma^{ij} p_{,i}
\end{eqnarray}
Furthermore, eqs.~(\ref{x1}) with the aid of eqs.~(\ref{x2})-(\ref{x3}) read
\begin{eqnarray}\label{x4}
&&D_{\tau} S^i+f_2\sigma^{ik}\gamma_{kl}\upsilon^l+\varepsilon a^i+\frac{1}{\sqrt{\gamma}}\frac{\partial}{\partial x^i}[\sqrt{\gamma} f_2\upsilon^i]
\Gamma_{kl}^i f_2\upsilon^k\upsilon^l+\gamma^{ij} p_{,j}\nonumber\\
&&+W^{ir}\gamma_{rr} a^r=\rho_e E^i+(\tilde{J}\times\tilde{B})^i
\end{eqnarray}
Taking the $r$-component of the last equation (\ref{x4}) we find
\begin{eqnarray}\label{x5}
&&D_{\tau} S^r+f_2\sigma^{r\phi}\gamma_{\phi\phi}\upsilon^{\phi}+\varepsilon a^r+\frac{1}{\sqrt{\gamma}}\frac{\partial}{\partial x^i}[\sqrt{\gamma} f_2\upsilon^r\upsilon^j]\nonumber\\
&&+\Gamma_{kl}^r f_2\upsilon^k\upsilon^l+\gamma^{rr} p_{,r}+\gamma_{rr} a^r[f_2\upsilon^r\upsilon^r+p\gamma^{rr}]=\rho_e E^r+(\tilde{J}\times\tilde{B})^r\nonumber\\
\end{eqnarray}
We re-write eq.~(\ref{x5}) as follows:
\begin{eqnarray}\label{x6}
&&D_{\tau} S^r+f_2\sigma^{r\phi}\gamma_{\phi\phi}\upsilon^{\phi}+[\varepsilon+p] a^r+\frac{1}{\sqrt{\gamma}}\frac{\partial}{\partial x^i}[\sqrt{\gamma} f_2\upsilon^r\upsilon^j]\nonumber\\
&&+\Gamma_{kl}^r f_2\upsilon^k\upsilon^l+\gamma^{rr} p_{,r}+\gamma_{rr} a^r[f_2\upsilon^r\upsilon^r]=\rho_e E^r+(\tilde{J}\times\tilde{B})^r\nonumber\\
\end{eqnarray}
Because of the form of $\varepsilon$ we find that
\begin{eqnarray}\label{x7}
p+\varepsilon&=&p+\Gamma^2(\rho+p\tilde{\upsilon}^2)p=p+\frac{1}{1-\tilde{\upsilon}^2}[\rho+p\tilde{\upsilon}^2]\nonumber\\
&=&\frac{\rho+p}{1-\tilde{\upsilon}^2}=f_2
\end{eqnarray}
Substitution of eq.~(\ref{x7}) into (\ref{x6}) we have
\begin{eqnarray}\label{x8}
&&a^r f_2=-\gamma^{rr} p_{,r}-D_{\tau} S^r-f_2\sigma^{r\phi}\gamma_{\phi\phi}\upsilon^{\phi}-\frac{1}{\sqrt{\gamma}}\frac{\partial}{\partial x^i}[\sqrt{\gamma} f_2\upsilon^r\upsilon^j]\nonumber\\
&&-\Gamma_{kl}^r f_2\upsilon^k\upsilon^l-\gamma_{rr} a^r f_2\upsilon^r\upsilon^r+\rho_e E^r+(\tilde{J}\times\tilde{B})^r
\end{eqnarray}
where
\begin{equation}\label{x9}
(\tilde{J}\times\tilde{B})^r=\varepsilon^{rkl}J_k B_l=\varepsilon^{r\phi\theta}J_{\phi} B_{\theta}=-\frac{B_{\theta} J_{\phi}}{\sqrt{\gamma}}
\end{equation}
From eqs.~(\ref{x8}) and (\ref{x9}) and the definition of the Fermi derivative $D_{\tau}$ we find
\begin{eqnarray}\label{x10}
a^r f_2[1+\gamma_{rr}\upsilon^r\upsilon^r]&=&-\gamma^{rr} p_{,r}-\frac{1}{\alpha}(n+im\omega) f_2\upsilon^r\nonumber\\
&-&\frac{1}{\alpha}[\Gamma_{t\phi}^r+\omega\Gamma_{\phi\phi}^r]f_2\upsilon^{\phi}-f_2\gamma_{\phi\phi}\upsilon^{\phi}\sigma^{r\phi}\nonumber\\
&-&\frac{1}{\sqrt{\gamma}}[(\sqrt{\gamma} f_2)_{,r}\upsilon^r\upsilon^r+f_2\sqrt{\gamma}(\upsilon^r\upsilon^r)_{,r}]\nonumber\\
&-&\Gamma_{kl}^r f_2\upsilon^k\upsilon^l+\rho_e E^r-\frac{B_{\theta} J_{\phi}}{\sqrt{\gamma}}
\end{eqnarray}
In the text, we consider  $\upsilon^i=(0,0,\upsilon^{\phi})$, and $\rho_e=0$. In this case, eqs.~(\ref{s9a}) and (\ref{x10}) simplify considerably and read
\begin{eqnarray}\label{x11}
\frac{(n+i m \omega)\delta\varepsilon}{\alpha}&=&-2\gamma_{rr} a^rf_2\delta\upsilon^r-\tilde{\upsilon}\cdot\tilde{\nabla}f_1-\delta\tilde{\upsilon}\cdot\tilde{\nabla}f_2\nonumber\\
&-&2\sigma_{r\phi}[f_2\upsilon^{\phi}\delta\upsilon^r)]-\gamma_{\phi\phi} J^{\phi}\delta E^{\phi}
\end{eqnarray}
\begin{eqnarray}\label{x12}
a^r f_2&=&-\gamma^{rr} p_{,r}-\frac{1}{\alpha}[\Gamma_{t\phi}^r+\omega\Gamma_{\phi\phi}^r+\alpha \Gamma_{\phi\phi}^r\upsilon^{\phi}]f_2\upsilon^{\phi}\nonumber\\
&-&f_2\gamma_{\phi\phi}\upsilon^{\phi}\sigma^{r\phi}-\frac{B_{\theta} J_{\phi}}{\sqrt{\gamma}}
\end{eqnarray}
We substitute eq.~(\ref{x12}) into eq.~(\ref{x11}) and find
\begin{eqnarray}\label{x13}
\frac{(n+i m \omega)\delta\varepsilon}{\alpha}&=&-2\gamma_{rr}\delta\upsilon^r\{-\gamma^{rr} p_{,r}\nonumber\\
&-&\frac{1}{\alpha}[\Gamma_{t\phi}^r+\omega\Gamma_{\phi\phi}^r+\alpha \Gamma_{\phi\phi}^r\upsilon^{\phi}]f_2\upsilon^{\phi}\nonumber\\
&-&f_2\gamma_{\phi\phi}\upsilon^{\phi}\sigma^{r\phi}-\frac{B_{\theta} J_{\phi}}{\sqrt{\gamma}}\}\nonumber\\
&-&\tilde{\upsilon}\cdot\tilde{\nabla}f_1-\delta\tilde{\upsilon}\cdot\tilde{\nabla}f_2\nonumber\\
&-&2\sigma_{r\phi}[f_2\upsilon^{\phi}\delta\upsilon^r)]-\gamma_{\phi\phi} J^{\phi}\delta E^{\phi}
\end{eqnarray}
Because of the relation $\tilde{\upsilon}\cdot\tilde{\nabla}f_1=\upsilon^{\phi} f_{1,\phi}$ and $\delta\tilde{\upsilon}\cdot\tilde{\nabla}f_2=\delta\upsilon^r f_{2,r} $, eq.~(\ref{x13}) reads
\begin{eqnarray}\label{x14}
\frac{(n+i m \omega)\delta\varepsilon}{\alpha}&=&-2\gamma_{rr}\delta\upsilon^r\{-\gamma^{rr} p_{,r}
\nonumber\\
&-&\frac{1}{\alpha}[\Gamma_{t\phi}^r+\omega\Gamma_{\phi\phi}^r+\alpha \Gamma_{\phi\phi}^r\upsilon^{\phi}]f_2\upsilon^{\phi}-\frac{B_{\theta} J_{\phi}}{\sqrt{\gamma}}\}\nonumber\\
&-&\upsilon^{\phi}f_{1,\phi}-\delta\upsilon^{r}f_{2,r}-\gamma_{\phi\phi} J^{\phi}\delta E^{\phi}
\end{eqnarray}
Furthermore, we re-write eqs.~(\ref{s5}) as
\begin{eqnarray}\label{x15}
f_1&\equiv&(\delta\rho+\delta p)\Gamma^2+(\rho+p)\delta\Gamma^2\nonumber\\
&=&\frac{(1+c_s^2)}{1-\tilde{\upsilon}^2}\delta \rho+2\frac{f_2}{1-\tilde{\upsilon}^2}\upsilon_k\delta\upsilon^k\nonumber\\
f_2&\equiv&(\rho+p)\Gamma^2=\frac{(1+c_s^2)}{1-\tilde{\upsilon}^2}\rho
\end{eqnarray}
and because of the form of the perturbations we find
\begin{eqnarray}\label{x16}
f_{2,\phi}&=&0\nonumber\\
f_{1,\phi}&=&\frac{(1+c_s^2)}{1-\tilde{\upsilon}^2}\delta \rho_{,\phi}+2\frac{f_2}{1-\tilde{\upsilon}^2}(\upsilon_k\delta\upsilon^k)_{,\phi}\nonumber\\
&=&im\{\frac{(1+c_s^2)}{1-\tilde{\upsilon}^2}\delta \rho+2\frac{f_2}{1-\tilde{\upsilon}^2}\upsilon_k\delta\upsilon^k\}=i m f_1
\end{eqnarray}
\begin{eqnarray}\label{x17}
f_{2,r}&=&(\rho+p)_{,r}\Gamma^2+(\rho+p)[\Gamma^2]_{,r}\nonumber\\
&=&\frac{\rho_{,r}+p_{,r}}{1-\tilde{\upsilon}^2}+f_2\frac{(\tilde{\upsilon}^2)_{,r}}{1-\tilde{\upsilon}^2}
\end{eqnarray}
From eqs.~(\ref{x14}), (\ref{x16}) and (\ref{x17}) we find
\begin{eqnarray}\label{x18}
\frac{(n+i m \omega)\delta\varepsilon}{\alpha}&=&-2\gamma_{rr}\delta\upsilon^r\{-\gamma^{rr} p_{,r}\nonumber\\
&-&\frac{1}{\alpha}[\Gamma_{t\phi}^r+\omega\Gamma_{\phi\phi}^r+\alpha \Gamma_{\phi\phi}^r\upsilon^{\phi}]f_2\upsilon^{\phi}\nonumber\\
&-&\frac{B_{\theta} J_{\phi}}{\sqrt{\gamma}}\}-i m \upsilon^{\phi}f_{1}-\delta\upsilon^{r}[\frac{\rho_{,r}+p_{,r}}{1-\tilde{\upsilon}^2}+f_2\frac{(\tilde{\upsilon}^2)_{,r}}{1-\tilde{\upsilon}^2}]\nonumber\\
&-&\gamma_{\phi\phi} J^{\phi}\delta E^{\phi}
\end{eqnarray}
Finally, for an isothermal equation of state (eq.~\ref{isothermal}), we re-write eq.~(\ref{x18}) as
\begin{eqnarray}\label{x22t}
\delta\varepsilon&=&-\frac{\alpha(n-i m\omega)}{n^2+m^2\omega^2}\{-2\upsilon^{\phi}(\frac{f_2}{\alpha})\delta\upsilon^r\gamma_{rr}[\Gamma_{t\phi}^r+\omega\Gamma_{\phi\phi}^r+\alpha\upsilon^{\phi}\Gamma_{\phi\phi}^r]\nonumber\\
&+&\delta\upsilon^{r}[\frac{\rho_{,r}}{1-\tilde{\upsilon}^2}(1-c_s^2+2c_s^2\tilde{\upsilon}^2)+f_2\frac{(\tilde{\upsilon}^2)_{,r}}{1-\tilde{\upsilon}^2}]+i m f_1\upsilon^{\phi}\nonumber\\
&-&\frac{2}{\sqrt{\gamma}}\gamma_{rr}B_{\theta}J_{\phi}\delta\upsilon^r+\gamma_{\phi\phi}J^{\phi}\delta E^{\phi}\}
\end{eqnarray}
which is equivalent to eq.~(\ref{v10b}).



\section*{Appendix~C}
We collect here complex expressions that are used in the main text of the paper.
\begin{eqnarray}\label{fin3a}
N_1&\equiv&(\frac{n}{m^2})\frac{A}{2\alpha r^4}[\frac{1+c_s^2}{1-\tilde{\upsilon}^2}][\frac{\tilde{\upsilon}^2}{1-\tilde{\upsilon}^2}]\rho(r^2\delta\upsilon^r)_{,r}\nonumber\\
&&-\frac{n}{m^2 \alpha}\frac{A}{r^4}[\frac{1+\tilde{\upsilon}^2}{1-\tilde{\upsilon}^2}f_2+\frac{B^2}{4\pi}](r^2\delta\upsilon^r)_{,r}-\frac{1}{4\pi}Re(B_{\theta}\delta B^{\theta})\nonumber\\
& + &\frac{n}{m \alpha^2}\frac{A}{r^2}\alpha\upsilon^{\phi}\frac{1}{4\pi}Re(iB_{\theta}\delta B^{\theta})
\end{eqnarray}

\begin{eqnarray}\label{fin3b}
N_2&\equiv&-\frac{A}{2 m\alpha r^4}[\frac{\tilde{\upsilon}^2\rho}{1-\tilde{\upsilon}^2}][G_6(r)(r^2\delta\upsilon^r)-m(\omega+2\alpha\upsilon^{\phi})(r^2\delta\upsilon^r)_{,r}]\nonumber\\
&&-\frac{A}{m\alpha r^4}[\Omega\frac{1+\tilde{\upsilon}^2}{1-\tilde{\upsilon}^2}f_2+\omega\frac{B^2}{4\pi}](r^2\delta\upsilon^r)_{,r}\nonumber\\
&+&\frac{A\upsilon^{\phi}}{mr^4}f_2[\frac{(\tilde{\upsilon}^2)_{,r}}{1-\tilde{\upsilon}^2}(r^2\delta\upsilon^r)-(r^2\delta\upsilon)_{,r}]\nonumber\\
&+&\frac{A}{mr^2}f_2(\upsilon^{\phi}\delta\upsilon^r)_{,r}+\frac{A}{m\alpha r^4}f_2 G_1(r)(r^2\delta\upsilon^r)\nonumber\\
&+&\frac{A}{m\alpha r^4}\frac{B^2}{4\pi}G_4(r)(r^2\delta\upsilon^r)-\frac{1}{4\pi}Im(B_{\theta}\delta B^{\theta})\nonumber\\
&+&\frac{\omega A\upsilon^{\phi}}{\alpha r^2}\frac{1}{4\pi} Im(iB_{\theta}\delta B^{\theta})
\end{eqnarray}
\begin{eqnarray}\label{fin8}
\Lambda_1&=&-(\frac{n}{\Delta\alpha})(f_2+\frac{B^2}{4\pi}) (r^2\delta\upsilon^r)\nonumber\\
&+&\frac{G_2(r)}{\alpha \Delta}[\frac{1+c_s^2}{1-\tilde{\upsilon}^2}](\alpha\upsilon^{\phi})\frac{n}{n^2+m^2\Omega^2}\nonumber\\
&\times&[\alpha\rho_{,r}(r^2\delta\upsilon^r)+\frac{\rho\upsilon_{\phi}G_6(r)}{2(1-\tilde{\upsilon}^2)}(r^2\delta\upsilon^r)-\frac{\rho\alpha \tilde{\upsilon}^2}{2(1-\tilde{\upsilon}^2)}(r^2\delta\upsilon^r)_{,r}]\nonumber\\
&+&\frac{n a^r}{\alpha \Delta}[\frac{\alpha c_s^2}{n^2+m^2\Omega^2}][\alpha\rho_{,r}(r^2\delta\upsilon^r)+\frac{\rho\upsilon_{\phi}G_6(r)}{2(1-\tilde{\upsilon}^2)}(r^2\delta\upsilon^r)\nonumber\\
&-& \frac{\rho\alpha\tilde{\upsilon}^2}{2(1-\tilde{\upsilon}^2)}(r^2\delta\upsilon^r)_{,r}]
\nonumber\\
&+&\frac{n\alpha a^r}{\Delta(n^2+m^2\omega^2)}\left\{\frac{1-c_s^2+2c_s^2\tilde{\upsilon}^2}{1-\tilde{\upsilon}^2}\rho_{,r}(r^2\delta\upsilon^r)\right.\nonumber\\
&-&\frac{2r^2f_2\upsilon^{\phi}}{\Delta\alpha}G_3(r)(r^2\delta\upsilon^r)+f_2\frac{(\tilde{\upsilon}^2)_{,r}}{1-\tilde{\upsilon}^2}(r^2\delta\upsilon^r)\nonumber\\
&-&[\frac{1+c_s^2}{1-\tilde{\upsilon}^2}][\frac{1}{n^2+m^2\Omega^2}]\nonumber\\
&\times&[m^2\Omega(\alpha\upsilon^{\phi}\rho_{,r}+\frac{\rho\tilde{\upsilon}^2 G_6(r)}{2(1-\tilde{\upsilon}^2)})(r^2\delta\upsilon^r)\nonumber\\
&-&\frac{\rho \tilde{\upsilon}^2}{2m(1-\tilde{\upsilon}^2)}(r^2\delta\upsilon^r)_{,r}(n^2+m^2\Omega\omega+2m^2\alpha\upsilon^{\phi}\Omega)]\nonumber\\
&-&\left.
\frac{2\tilde{\upsilon}^2}{1-\tilde{\upsilon}^2}f_2(r^2\delta\upsilon^r)_{,r}
-\frac{3}{4\pi}[B_{\theta} B_{,r}^{\theta}+(\frac{2}{r}+a^r) B^2](r^2\delta\upsilon^r)]\right\}\nonumber\\
&-&\frac{\alpha a^r}{\Delta}\frac{1+c_s^2}{1-\tilde{\upsilon}^2}\frac{m^2\omega}{(n^2+m^2\omega^2)(n^2+m^2\Omega^2)}\nonumber\\
&\times&[(\alpha\upsilon^{\phi}\rho_{,r}+\frac{\rho\tilde{\upsilon}^2}{2(1-\tilde{\upsilon}^2)}G_{6}(r))(r^2\delta\upsilon^r)-\frac{\alpha\upsilon^{\phi}\rho\tilde{\upsilon}^2}{2(1-\tilde{\upsilon}^2)}(r^2\delta\upsilon^r)_{,r}]\nonumber\\
&+&\frac{2n\alpha a_r}{4\pi r^2} \frac{B_{\theta} B_{,r}^{\theta}}{n^2+m^2\Omega^2}(r^2\delta\upsilon^r)\nonumber\\
&+&\frac{n}{4\pi\alpha\Delta}\frac{B_{\theta} B_{,r}^{\theta}}{n^2+m^2\Omega^2}\alpha\upsilon^{\phi}G_5(r)(r^2\delta\upsilon^r)
\end{eqnarray}
\begin{eqnarray}\label{fin9}
\Lambda_2 &=&-\frac{m}{\Delta\alpha}(f_2\Omega+\omega\frac{B^2}{4\pi})(r^2\delta\upsilon^r)-\frac{f_2}{m\Delta}\Gamma_{\phi\phi}^r \upsilon^{\phi}(r^2\delta\upsilon^r)_{,r}\nonumber\\
&+& \frac{ G_2(r)}{\Delta\alpha}[\frac{1+c_s^2}{1-\tilde{\upsilon}^2}]\frac{(\alpha\upsilon^{\phi})}{n^2+m^2\Omega^2}[m\Omega(\alpha\rho_{,r}+\frac{\rho\Gamma^2\upsilon_{\phi}G_6(r)}{2})(r^2\delta\upsilon^r)\nonumber\\
&&-\frac{\rho\Gamma^2\upsilon_{\phi}}{2m}(r^2\delta\upsilon^r)_{,r}(n^2+m^2\omega\Omega+2m^2\Omega\alpha\upsilon^{\phi})]\nonumber\\
&-&(\frac{1}{m})\frac{G_2(r)}{\alpha\Delta}[\frac{1+\tilde{\upsilon}^2}{1-\tilde{\upsilon}^2}]f_2(r^2\delta\upsilon^r)_{,r}\nonumber\\
&-&(\frac{a^r}{\alpha\Delta})\frac{\alpha c_s^2}{n^2+m^2\Omega^2}[m\Omega(\alpha\rho_{,r}+\frac{\rho\upsilon_{\phi}\Gamma^2G_6(r)}{2})(r^2\delta\upsilon^r)\nonumber\\
&&-\frac{\rho\upsilon_{\phi}\Gamma^2}{2m}(r^2\delta\upsilon^r)_{,r}(n^2+m^2\omega\Omega+2m^2\Omega\alpha\upsilon^{\phi})]\nonumber\\
&-&\frac{m\omega}{n^2+m^2\omega^2}\frac{\alpha a^r}{\Delta} \left\{\frac{1-c_s^2+2c_s^2\tilde{\upsilon}^2}{1-\tilde{\upsilon}^2}\rho_{,r}(r^2\delta\upsilon^r)\right.\nonumber\\
&-&\frac{2r^2 f_2}{\Delta\alpha}\upsilon^{\phi}G_3(r)(r^2\delta\upsilon^r)+f_2\frac{(\tilde{\upsilon}^2)_{,r}}{1-\tilde{\upsilon}^2}(r^2\delta\upsilon^r)\nonumber\\
&-&[\frac{1+c_s^2}{1-\tilde{\upsilon}^2}]\frac{m}{n^2+m^2\Omega^2}[m\Omega(\alpha\rho_{,r}+\frac{\rho\tilde{\upsilon}^2\Gamma^2G_6(r)}{2})(r^2\delta\upsilon^r)\nonumber\\
&&-\frac{\rho\tilde{\upsilon}^2\Gamma^2}{2m}(r^2\delta\upsilon^r)_{,r}(n^2+m^2\omega\Omega+2m^2\Omega\alpha\upsilon^{\phi})]\nonumber\\
&-&\left.
2\frac{\tilde{\upsilon}^2}{1-\tilde{\upsilon}^2}f_2(r^2\delta\upsilon^r)_{,r}
-\frac{3}{4\pi}[B_{\theta} B_{,r}^{\theta}+(\frac{2}{r}+a^r) B^2](r^2\delta\upsilon^r)\right\}\nonumber\\
&&-(\frac{\alpha a^r}{\Delta})\frac{1+c_s^2}{1-\tilde{\upsilon}^2}\frac{m n}{(n^2+m^2\omega^2)(n^2+m^2\Omega^2)}\nonumber\\
&\times&[(\alpha\upsilon^{\phi}\rho_{,r}+\frac{\rho\Gamma^2\tilde{\upsilon}^2G_6(r)}{2})(r^2\delta\upsilon^r)-\frac{\rho\Gamma^2\upsilon^{\phi}\tilde{\upsilon}^2}{2}\alpha(r^2\delta\upsilon^r)_{,r}]\nonumber\\
&-&\frac{m\Omega\alpha}{4\pi r^2}\frac{B_{\theta} B_{,r}^{\theta}}{n^2+m^2\Omega^2}(r^2\delta\upsilon^r)[2a^r+r^2G_5(r)\frac{\upsilon^{\phi}}{\alpha\Delta}]\nonumber\\
&+&\frac{1}{\alpha\Delta m}\frac{B^2}{4\pi}(r^2\delta\upsilon^r)_{,r}G_5(r)\nonumber\\
\end{eqnarray}
and
\begin{eqnarray}\label{fin6a}
G_1(r)&\equiv&(\frac{2}{r}+\Gamma_{\phi r}^{\phi})\alpha\upsilon^{\phi}+\Gamma_{t r}^{\phi}+\omega\Gamma_{\phi r}^{\phi}-\omega a_r+\alpha \gamma_{rr}\sigma^{r\phi}\nonumber\\
G_2(r)&\equiv&\Gamma_{t\phi}^r+\Gamma_{\phi\phi}^r \Omega+\alpha\gamma_{\phi\phi}\sigma^{r\phi}\nonumber\\
G_3(r)&\equiv&\Gamma_{t\phi}^r+\Gamma_{\phi\phi}^r \Omega\nonumber\\
G_4(r)&\equiv&\Gamma_{t r}^{\phi}+\omega\Gamma_{\phi r}^{\phi}+\alpha\gamma^{\phi\phi}\sigma_{r\phi}\nonumber\\
G_5(r)&\equiv&\Gamma_{t\phi}^r+\Gamma_{\phi\phi}^r \omega+\alpha\gamma^{rr}\sigma_{r\phi}\nonumber\\
G_6(r)&\equiv& \Gamma_{t r}^{\phi}+\omega\Gamma_{\phi r}^\phi-\omega a_r
\end{eqnarray}

\section*{Appendix~D}
The differential equation (\ref{dif4}) admits the solutions
\begin{eqnarray}\label{dif5}
w_2&\equiv& c_2 r^{\frac{1-\xi}{2}}H_2,~~\mbox{~for~}~~r>r_{\rm ISCO}\ ,\nonumber\\
w_1&\equiv& c_1r^{\frac{1+\xi}{2}}H_1,~~\mbox{~for~}~~r<r_{\rm ISCO}\ ,
\end{eqnarray}
where
\begin{eqnarray}\label{c1}
\xi&\equiv&\sqrt{1+4m^2}\nonumber\\
H_1&\equiv&HeunB [-\xi, k_2,k_3,k_4,\frac{k_5}{r}]\nonumber\\
H_2&\equiv&HeunB [-\xi,k_2,k_3,k_4,\frac{k_5}{r}]\nonumber\\
k_2&=&-\frac{\sqrt{10}}{5}\sqrt{\frac{7(1+c_s^2)+2u_A^2}{1+c_s^2}}\nonumber\\
k_3&=&3\nonumber\\
k_4&=&\frac{2\sqrt{10}}{5}\frac{[7(1+c_s^2)+2u_A^2+m^2(1+c_s^2+4u_A^2)]}{\sqrt{(1+c_s^2)[7(1+c_s^2)+2u_A^2]}}\nonumber\\
k_5&=&\frac{M\sqrt{10}}{4}\frac{\sqrt{(1+c_s^2)[7(1+c_s^2)+2u_A^2]}}{1+c_s^2+u_A^2}
\end{eqnarray}
Here, $c_1$ is an arbitrary constant, and $c_2=c_1 r^{\,\xi} H_1/H_2|_{r=r_{\rm ISCO}}$ is chosen such as to guarantee the continuity of $w(r)$ at the interface at $r=r_{\rm ISCO}$. $u_A^2\equiv B^2/(4\pi\rho)$ and $HeunB [\xi,k_2,k_3,k_4,k_5/r]$ is the Heun Biconfluent function which is the solution of the Heun Biconfluent equation~\citep{n1}.

\section*{Appendix~E}
Here we present the explicit expressions $R$, $L_1$, $L_2$, $\tilde{L}_1$ and $\tilde{L}_2$ of eq.~(\ref{master1}).
$R\equiv R_1+\tilde{R}_1$, where
\begin{eqnarray}\label{master2}
R_1 &\equiv&\frac{1+\tilde{\upsilon}^2}{(1-\tilde{\upsilon}^2)^2}(1+c_s^2){\cal D}\{\rho\}+{\cal D}\{\frac{B^2}{4\pi}\}\nonumber\\
&-&\xi[\frac{1+\tilde{\upsilon}^2}{(1-\tilde{\upsilon}^2)^2}(1+c_s^2){\cal P}\{\rho\}+{\cal P}\{\frac{B^2}{4\pi}\}]\nonumber\\
&-&(\frac{k_5 k_2}{r})[\frac{1+\tilde{\upsilon}^2}{(1-\tilde{\upsilon}^2)^2}(1+c_s^2){\cal D}\{\rho\}+{\cal D}\{\frac{B^2}{4\pi}\}]\nonumber\\
&-&(\frac{k_5 k_4}{4 m^2 r})[\frac{1+\tilde{\upsilon}^2}{(1-\tilde{\upsilon}^2)^2}(1+c_s^2){\cal P}\{\rho\}+{\cal P}\{\frac{B^2}{4\pi}\}]\nonumber\\
\label{master2a}
\tilde{R}_1 &\equiv&-\frac{\tilde{\upsilon}^2(1+c_s^2)}{2(1-\tilde{\upsilon}^2)^2}[{\cal D}\{\rho\}-\xi{\cal P}\{\rho\}\nonumber\\
&+&[\frac{\tilde{\upsilon}^2(1+c_s^2)}{2(1-\tilde{\upsilon}^2)^2}][\frac{k_5 k_2}{r}]{\cal D}\{\rho\}\nonumber\\
&-&[\frac{\tilde{\upsilon}^2(1+c_s^2)}{2(1-\tilde{\upsilon}^2)^2}][\frac{k_5 k_4}{4m^2r}][{\cal D}\{\rho\}-\xi{\cal P}\{\rho\}]\nonumber\\
\label{master3}
L_1&\equiv&(\frac{2r^5\alpha^2a^r}{A\Delta })\{\frac{1-c_s^2+2\tilde{\upsilon}^2c_s^2}{1-\tilde{\upsilon}^2}{\cal D}\{\rho\}-3{\cal D}\{\frac{B^2}{8\pi}\}\}\nonumber\\
\tilde{L}_1&\equiv&-(\frac{2r^5\alpha^2a^r}{A\Delta })[\frac{1+c_s^2}{1-\tilde{\upsilon}^2}](\alpha\upsilon^{\phi})(2\omega+\alpha\upsilon^{\phi}){\cal D}\{\rho\}\nonumber\\
\label{master4}
L_2&\equiv&(\frac{2r^3}{A}){\cal D}\{\frac{B^2}{8\pi}\}\{2\alpha^2 a_r\nonumber\\
&+&\alpha[-a_r(\frac{A\upsilon^{\phi}}{r^2})+(\frac{A\upsilon^{\phi}}{r^2})_{,r}]\Omega+(\frac{r^2\alpha\upsilon^{\phi}}{\Delta})G_5(r)\}\nonumber\\
\tilde{L}_2&\equiv&(\frac{2r^5\alpha a^r}{A\Delta })(\alpha c_s^2){\cal D}\{\rho\}+(\frac{2\alpha G_2(r) r^5}{A \Delta})[\frac{1+c_s^2}{1-\tilde{\upsilon}^2}](\alpha\upsilon^{\phi}){\cal D}\{\rho\}\nonumber\\
\end{eqnarray}
All of the above expressions are evaluated at the interface at $r=r_{\rm ISCO}$. We remind the reader that we have defined here ${\cal D} \{f\}\equiv f_{(2)}-f_{(1)}$ and ${\cal P} \{f\}\equiv f_{(2)}+f_{(1)}$ at the ISCO.

\end{document}